\documentclass[%
 reprint,
 superscriptaddress,
nofootinbib,
 amsmath,amssymb,
 aps, 
 prl,
]{revtex4-2}

\usepackage{graphicx}% Include figure files
\usepackage{dcolumn}% Align table columns on decimal point
\usepackage{bm}% bold math
\usepackage{color}
\usepackage{physics}
\usepackage{dsfont}
\usepackage{mathtools}
\usepackage{lipsum}
\usepackage[top=2cm, bottom=2cm, left=1.5cm, right=1.5cm]{geometry}
\usepackage[svgnames]{xcolor}
\usepackage{hyperref}% add hypertext capabilities
\hypersetup{
    colorlinks=true,
    citecolor=blue,
    linkcolor=blue,
    filecolor=blue,  
    urlcolor=DarkBlue,
    pdftitle={Deterministic Mechanical Wigner Negativity via Nonlinear Cavity Quantum Optomechanics},
    pdfpagemode=FullScreen,
    }
\usepackage[mathlines]{lineno}% Enable numbering of text and display math
\usepackage{orcidlink}

\newcommand{\rme}{\mathrm{e}}
\newcommand{\rmi}{\mathrm{i}}
\newcommand{\rmd}{\mathrm{d}}

\newcommand{\nl}{n_{{l}}}

\newcommand{\wm}{\omega_{{m}}}
\newcommand{\wc}{\omega_{{c}}}
\newcommand{\wl}{\omega_{{l}}}
\newcommand{\ain}{a_{{in}}}
\newcommand{\aout}{a_{{out}}}
\newcommand{\alphain}{\alpha_{{in}}}
\newcommand{\rin}{r_{{in}}}

\newcommand{\rhoin}{\rho_{{i}}}
\newcommand{\rhof}{\rho}

\newcommand{\Ph}{\mathcal{P}}

\newcommand{\Db}{\bar{\Delta}}
\newcommand{\tauth}{\tau_{\mathrm{th}}}
\newcommand{\bd}{b^{\dagger}}

\begin{document}    
\preprint{APS/123-QED}

\title{Deterministic Mechanical Wigner Negativity via Nonlinear Cavity Quantum Optomechanics in the Unresolved-Sideband Regime}

\author{Jack Clarke\,\orcidlink{0000-0001-8055-449X}}
 \email{jack-clarke@ucl.ac.uk}
 \affiliation{Quantum Measurement Lab, Blackett Laboratory, \href{https://ror.org/041kmwe10}{Imperial College London}, London SW7 2BW, United Kingdom}
\affiliation{Department of Physics and Astronomy, \href{https://ror.org/02jx3x895}{University College London}, Gower Street, WC1E 6BT London, United Kingdom}
\author{Pascal Neveu\,\orcidlink{0000-0002-7671-2763}}
\affiliation{Center for Nanophotonics, \href{https://ror.org/038x9td67}{AMOLF}, Science Park 104, 1098 XG Amsterdam, The Netherlands}%
\author{Ewold Verhagen\,\orcidlink{0000-0002-0276-8430}}
\email{verhagen@amolf.nl}
\affiliation{Center for Nanophotonics, \href{https://ror.org/038x9td67}{AMOLF}, Science Park 104, 1098 XG Amsterdam, The Netherlands}%
\author{Michael R.~Vanner\,\orcidlink{0000-0001-9816-5994}}
\email{m.vanner@imperial.ac.uk}
\homepage{www.qmeas.net}
\affiliation{Quantum Measurement Lab, Blackett Laboratory, \href{https://ror.org/041kmwe10}{Imperial College London}, London SW7 2BW, United Kingdom}

\date{\today}
\begin{abstract}
%\lipsum[1]
%
Non-Gaussian quantum states of mechanical motion exhibiting Wigner negativity offer promising capabilities for quantum technologies and tests of fundamental physics.
Within the field of cavity quantum optomechanics, deterministic preparation of nonclassical mechanical states with such Wigner negativity is a highly sought goal but is challenging as the intracavity interaction Hamiltonian is linear in mechanical position.
Here, we show that, despite this form of interaction, by utilizing the nonlinearity of the cavity response with mechanical position, mechanical Wigner negativity can be prepared deterministically in the unresolved-sideband regime, without additional nonlinearities, nonclassical drives, or conditional measurements.
In particular, we find that Wigner negativity can be prepared with an optical pulse, even without single-photon strong coupling, and the negativity persists in the steady state of a continuously driven system. 
Our results deepen our understanding of cavity-enhanced radiation pressure and establish a pathway for deterministic preparation of nonclassical mechanical states in the unresolved sideband regime.
\end{abstract}
%

\maketitle

%%%%%%%%%%%%%%%%%%%%%%%%%%%%%%%%%%%%%%%%%%%%%%%%%%%%%%%%%%%%%%%%%
\textit{Introduction.}---A particularly distinguishing property of nonclassical states is their ability to exhibit negativity in their phase-space quasiprobability distributions.
This phase-space negativity manifests in quasiprobability distributions such as the Glauber-Sudarshan $P$ function~\cite{sudarshan1963equivalence,glauber1963coherent} and the Wigner distribution~\cite{wigner1932quantum}, and has been observed across a variety of platforms, ranging from optical~\cite{lvovsky2001quantum, zavatta2004quantum, yoshikawa2013creation} and microwave fields~\cite{deleglise2008reconstruction,hofheinz2009synthesizing}, to trapped-ions~\cite{leibfried1996experimental}, atomic ensembles~\cite{mcconnell2015entanglement}, and mechanical modes coupled to superconducting qubits~\cite{satzinger2018quantum, chu2018creation}. Beyond its foundational significance, such negativity enables a broad range of applications, including quantum speedup in computation~\cite{veitch2012negative,mari2012positive}, and tests of fundamental physics~\cite{bose1999scheme,marshall2003towards}.

By utilizing the radiation-pressure interaction, cavity quantum optomechanics provides a powerful toolset for motional quantum state engineering, and while significant experimental progress has been made within the linearized approximation~\cite{aspelmeyer2014cavity}, this regime cannot generate Wigner negativity from Gaussian input states deterministically.
Therefore, to generate mechanical negativity within the linearized approximation, additional nonlinearities are required, which include, for example, heralding based on photon counting~\cite{borkje2011proposal,vanner2013quantum}, state transfer of non-Gaussian optical states~\cite{mancini2003scheme,marek2010noise,bennett2016quantum}, and coupling to two-level systems~\cite{arcizet2011single}. 
Moreover, even beyond the linearized approximation, many schemes for generating mechanical negativity rely on such conditional measurements~\cite{vacanti2008optomechanical, hoff2016measurement, ringbauer2018generation, clarke2018growing, neumeier2018exploring}. 
By contrast, driving a nonlinear cavity optomechanical system in the resolved-sideband regime, where the mechanical frequency exceeds the cavity linewidth, mechanical Wigner negativity can be generated deterministically~\cite{rabl2011photon,nunnenkamp2011single,qian2012quantum,nation2013nonclassical,lorch2014laser,wise2024nonclassical}.
Recent theoretical research in this regime has focused on how this mechanical negativity may be enhanced by utilizing a variety of techniques including reservoir engineering, additional mechanical anharmonicities, and multiple optical drives~\cite{rips2012steady, tan2013generation, brunelli2018unconditional, hauer2023nonlinear}. In addition to resolved-sideband operation, schemes to create nonclassical states of motion in the unresolved-sideband regime, where the cavity linewidth exceeds the mechanical frequency, are of particular interest for a wide range of applications including sensing~\cite{aspelmeyer2014cavity}, and investigating decoherence and links between quantum mechanics and gravity~\cite{bassi2013models, kanari2021can, bose2023massive}.
This regime also offers a promising route to achieving strong single-photon nonlinearities, which is a highly sought goal within optomechanics~\cite{purdy2010tunable, xuereb2012strong, leijssen2017nonlinear, neumeier2018exploring} and also superconducting analogues of the radiation-pressure interaction~\cite{bothner2021photon}.
Moreover, with a growing number of unresolved-sideband experiments now observing strong nonlinearities~\cite{brennecke2008cavity, purdy2010tunable, brawley2016nonlinear, leijssen2017nonlinear, muhonen2019state, fedorov2020thermal, pluchar2023thermal}, deterministic preparation of mechanical Wigner negativity in the unresolved-sideband regime would open a rich avenue for research and development.

In this Letter, we show that mechanical Wigner negativity can be prepared in the unresolved-sideband regime via \emph{solely} the cavity-enhanced radiation-pressure interaction. Our approach utilizes both radiation-pressure and, crucially, the nonlinear response of the cavity with mechanical position and does not require any additional nonlinearities, conditional measurements, single-photon strong coupling, strong or nonclassical drives, or ground-state cooling. 
To generate mechanical negativity, even for weak single-photon coupling, a pulsed optical input may be utilized, and to enhance the negativity we propose a combination of drive detuning, and optical and mechanical squeezing. Indeed, various proposals have recently considered the use of squeezing to amplify quantum fluctuations in order to enhance the effects of other nonlinearities~\cite{vanner2011selective, roda2024macroscopic, marti2024quantum, rosiek2024quadrature}. 
Additionally, we find that for a continuous drive, mechanical negativity persists in the steady state for single-photon strong coupling. To demonstrate the robustness of our scheme, we explore how open-quantum-system dynamics affect the generation of mechanical Wigner negativity and we analyze how photon counting can be used to enhance the negativity. These results demonstrate that utilizing the nonlinear cavity response provides a powerful approach for deterministic mechanical quantum state engineering, widening the scope for experiments in the unresolved sideband regime.
%

%%%%%%%%%%%%%%%%%%%%%%%%%%%%%%%%%%%%%%%%%%%%%%%%%%%%%%%%%%%%%%%%%
\textit{Nonlinear cavity quantum optomechanics.}---The intracavity radiation-pressure interaction is described by $H/\hbar=-{g}_{0}{a}^{\dagger}a(b+b^{\dagger})$, where $g_{0}$ is the optomechanical coupling rate, and $a$ ($b$) is the annihilation operator of the optical (mechanical) mode. By solving the optomechanical Langevin equations and the optical input-output relation for an adiabatic cavity where $\dot{a}\simeq0$, one obtains the nonlinear unitary $U=\rme^{\rmi\varphi(X)\nl}$, where $\nl$ is the photon number operator of the input optical mode defined over the duration of a short optical pulse or a small time increment for a continuous drive, $\varphi(X)=\arg(f)$ is the optical phase and $f(X)=\left[1+\rmi\left(\frac{\mu}{2}{X}+\Db\right)\right]/\left[1-\rmi\left(\frac{\mu}{2}{X}+\Db\right)\right]$ is the nonlinear cavity response function, which relates the input and output optical fields via $a_{out}(t)=f(X)\,a_{in}(t)$. Here, $X = (b+b^{\dagger})/\sqrt{2}$ and $P = -\rmi(b-b^{\dagger})/\sqrt{2}$ are the dimensionless mechanical position and momentum quadratures, respectively, $\mu=\sqrt{8}g_{0}/\kappa$ is the nonlinear coupling strength, $\kappa$ is the cavity {amplitude} decay rate, and $\Db$ is the mean drive detuning normalized in units of $\kappa$ from the cavity's resonance at zero mechanical displacement. 
The nonlinear cavity response function $f(X)$ can be used to simply describe how the amplitude and phase of light respond to the cavity, and the form of this complex function is well known throughout cavity-based optics research. Recently, some of the consequences of $f(X)$ have been explored in cavity optomechanics~\cite{aspelmeyer2014cavity, neumeier2018exploring, clarke2023cavity}, and we’d like to highlight that there is rich optomechanics physics that results from this response that remains to be explored.
%

%%%%%%%%%%%%%%%%%%%%%%%%%%%%%%%%%%%%%%%%%%%%%%%%%%%%%%%%%%%%%%%%%
%\textit{Arrowhead states and pulsed operation.}
\textit{Mechanical negativity via pulsed optomechanics.}---Pulsed quantum optomechanics~\cite{vanner2011pulsed} utilizes pulsed interactions much shorter than a mechanical period $2\pi/\wm$, requiring operation in the unresolved-sideband regime $\kappa\gg\wm$. This regime ensures free mechanical evolution and dissipation are negligible over the optomechanical interaction. Following a pulsed optomechanical interaction, if no measurement is made on the optical output, the mechanical state $\rhof$ is given by the deterministic mapping $\rhof=\tr_{l}\left(U\rhoin\otimes\ket{\psi}\bra{\psi} U^{\dagger}\right)$ where $\ket{\psi}=\sum_{n=0}^{\infty}c_{n}\ket{n}$ is a general optical input state, $\rhoin$ is the initial mechanical state, and the subscript $l$ indicates the partial trace is taken over the optical subspace.
Given the Wigner function of this initial mechanical state $W_{i}(X,P)=\frac{1}{2\pi}\int_{-\infty}^{+\infty}\rme^{\rmi Pu}\bra{X-u/2}\rhoin\ket{X+u/2}\rmd{u}$, the Wigner function of the final mechanical state is $W(X,P)=\frac{1}{2\pi}\int_{-\infty}^{+\infty}\rme^{\rmi Pu}\mathcal{K}(X,u)\bra{X-u/2}\rhoin\ket{X+u/2}\rmd{u}$, where the kernel is $\mathcal{K}(X,u) = \sum_{n=0}^{\infty}|c_{n}|^2\left[f(X-u/2)f^{*}(X+u/2)\right]^n$.

To see how our scheme deterministically generates mechanical Wigner negativity, let's first re-examine Hudson's theorem~\cite{hudson1974wigner,mandilara2009extending}, which states that a necessary and sufficient condition for Wigner negativity is that the wavefunction of a pure state be the exponential of a polynomial beyond quadratic order. 
Thus, terms beyond quadratic order in the expansion of $\varphi(X)$ are required to generate a non-Gaussian quantum state exhibiting Wigner negativity from a Gaussian initial state.
Crucially, such terms cannot be generated from the interaction Hamiltonian alone and when only the radiation-pressure nonlinearity is considered $\varphi(X) \rightarrow {\mu X+2\Db}$~\cite{pikovski2012probing,wang2017enhancing}. In this case, the mechanical state reduces to a classical mixture of displaced initial states, $\sum_{n=0}^{\infty}|c_{n}|^2 W_{i}(X,P-n\mu)$, which cannot produce negativity even when $g_0/\kappa>1$. By contrast, when the response of the cavity with mechanical position is included, $\varphi(X)$ has higher-order contributions that lead to the generation of Wigner negativity. Physically, the resultant negativity may be more intuitively understood by noting that, for a resonant optical pulse ($\Db = 0$), the momentum transfer is strongest for mechanical positions close to zero and is weaker for non-zero positions owing to the Lorentzian dependence of the intracavity mean photon number with mechanical displacement. 
This emergence of deterministic Wigner negativity in the unresolved-sideband regime was not previously predicted and highlights the importance of the role of optical input-output and the subtle nonlinear nature of the cavity response.
%

%%%%%%%%%%%%%%%%%%%%%%%%%%%%%%%%%%%%%%%%%%%%%%%%%%%%%%%%%%%%%%%%%
%%%%%%%%%%%%%%%%%%%%%%%%%%%% Figure 1 %%%%%%%%%%%%%%%%%%%%%%%%%%%
%%%%%%%%%%%%%%%%%%%%%%%%%%%%%%%%%%%%%%%%%%%%%%%%%%%%%%%%%%%%%%%%%
\begin{figure*}
    \centering
    \includegraphics[width=\textwidth]{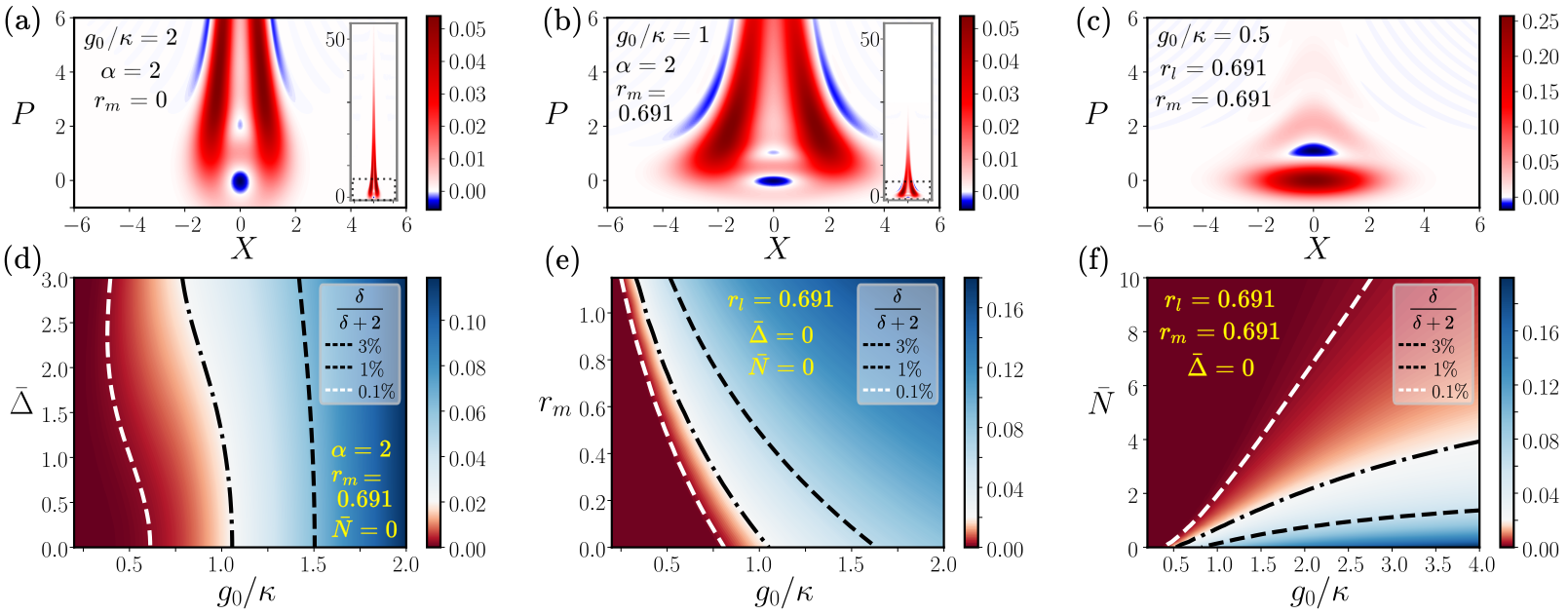}
    \caption{Deterministic mechanical Wigner negativity via pulsed nonlinear cavity optomechanics. (a--c) Mechanical Wigner functions $W(X,P)$ and (d--f) the negative volume indicator $\delta$. Here, the relevant parameters are $g_{0}/\kappa$, the amplitude of the coherent pulse $\alpha$, the optical $r_{l}$ and mechanical $r_{m}$ squeezing parameters, the normalized optical detuning $\bar{\Delta}$, and the initial thermal mechanical occupation $\bar{N}$. (a) A mechanical state generated via the pulsed interaction between an initial mechanical ground state and a coherent pulse of light with amplitude $\alpha=2$, $\Db=0$, and $g_{0}/\kappa=2$. (b) Prior to the pulsed optomechanical interaction, a squeezing operation may be applied to the initial mechanical state to enhance mechanical negativity. Here, the initial ground state is squeezed by $6\,\mathrm{dB}$ ($r_{m}=0.691$) in momentum and $g_{0}/\kappa=1$. (c) Utilizing an optical squeezed vacuum state assists generating Wigner negativity. In this plot, the optical squeezing parameter is $r_{l}=0.691$ and $g_{0}/\kappa=0.5$. To aid comparison, the above parameters are chosen so that Wigner functions (a--c) all possess $\delta=0.016$, demonstrating how optical and mechanical squeezing can enhance Wigner negativity generation even when $g_{0}/\kappa<1$. (d) $\delta$ plotted as a function of $\bar{\Delta}$ and $g_{0}/\kappa$, after a coherent pulse of light interacts with the mechanical mode, illustrating how detuning can also enhance negativity, especially at smaller values $g_{0}/\kappa$. The dashed curves indicate the boundaries of percentages of the negative volume to the positive volume. (e) $\delta$ plotted as a function of $r_{m}$ and $g_{0}/\kappa$ following a pulsed optomechanical interaction with a squeezed vacuum input pulse, which illustrates how both mechanical and optical squeezing may be utilized as a tool to further increase Wigner negativity. (f) $\delta$ plotted as a function of $\bar{N}$ and $g_{0}/\kappa$. Notably, even at finite values of initial thermal occupation, Wigner negativity is still be deterministically generated.
    }
    \label{fig_1_main}
\end{figure*}
%%%%%%%%%%%%%%%%%%%%%%%%%%%%%%%%%%%%%%%%%%%%%%%%%%%%%%%%%%%%%%%%%
%
%%%%%%%%%%%%%%%%%%%%%%%%%%%%%%%%%%%%%%%%%%%%%%%%%%%%%%%%%%%%%%%%%

In Fig.~\ref{fig_1_main}, we plot Wigner functions of the mechanical states generated deterministically via pulsed nonlinear cavity optomechanics. To quantify the negativity of the mechanical Wigner functions we evaluate the total negative volume indicator $\delta=\int|W(X,P)|\rmd{X}\rmd{P}-1$~\cite{kenfack2004negativity}, which is a measure of nonclassicality~\cite{tan2020negativity} and genuine non-Gaussianity~\cite{albarelli2018resource}.
In Fig.~\ref{fig_1_main}(a), mechanical Wigner negativity is generated by driving the cavity with a coherent pulse of light of amplitude $\alpha$. 
Here, the mechanics is initially in the ground state, $\alpha=2$, $\Db=0$, and $g_{0}/\kappa=2$, which generates $\delta=0.016$. The mechanical Wigner function resembles an \emph{arrowhead} shape as the momentum kick per photon is maximum at $X=0$ and tends to zero at large values of $X$. Fig.~\ref{fig_1_main}(b) demonstrates how momentum squeezing the initial mechanical state reduces the value of $g_{0}/\kappa$ needed to generate negativity. In this example, $6\,\mathrm{dB}$ of mechanical squeezing (a squeezing parameter $r_{m}=0.691$) enables the same value of $\delta$ to be realized as in Fig.~\ref{fig_1_main}(a), even when $g_{0}/\kappa$ is halved to $1$. Experimentally, this mechanical squeezing could be achieved probabilistically, via pulsed interaction and measurement~\cite{vanner2011pulsed, clarke2023cavity}, or deterministically, via a sequence of pulses~\cite{khosla2013quantum} or with measurement and feedback~\cite{mashaal2024strong}. Fig.~\ref{fig_1_main}(c) shows the mechanical state generated when the cavity is driven with a pulse of light in a squeezed vacuum state, while the mechanical mode is also squeezed by the same amount. In this plot, $6\,\mathrm{dB}$ of optical squeezing (along any quadrature~\cite{supp}) enables the same value of $\delta$ to be achieved as in the previous plots, but with a further reduced value of $g_{0}/\kappa=0.5$. Thus, utilizing a combination of optical and mechanical squeezing, mechanical negativity can be generated without requiring single-photon strong coupling, i.e. with $g_{0}/\kappa<1$.

In Figs~\ref{fig_1_main}(d-f) we explore the dependence on the negative volume with system parameters. In Fig.~\ref{fig_1_main}(d), we plot $\delta$ as a function of $g_{0}/\kappa$ and $\Db$ in the absence of squeezing. Here, a stronger dependence on $\Db$ is observed for low values of $g_{0}/\kappa$. This dependence is exemplified by the dashed curves, which mark the boundaries where the total negative volume $\delta/2$ is equal to $0.1\%$, $1\%$, and $3\%$ of the total positive volume. % 
For $\Db=0$ the mechanical states are symmetric about $X=0$, cf. Figs~\ref{fig_1_main}(a--c), while introducing a non-zero detuning can bias the Wigner negativity to one side of the distribution, which can increase Wigner negativity~\cite{supp}. Notably, the Wigner negativity of states with lower $g_{0}/\kappa$ values are more sensitive to this redistribution of phase-space volume. Furthermore, for a given value of $\Db$, the absolute value of the minimum of the Wigner function also increases with $g_{0}/\kappa$.
In Fig.~\ref{fig_1_main}(e), the dependence of $\delta$ on the mechanical squeezing parameter $r_{m}$ is investigated. This contour plot shows that increasing $r_{m}$ provides a route towards generating Wigner negativity in systems with $g_{0}/\kappa$ significantly smaller than unity. 
Despite the pulsed interaction occurring over a timescale much shorter than the open-system dynamics, any initial mechanical thermal occupation will still limit the nonclassicality generated. Thus, Fig.~\ref{fig_1_main}(e) investigates the robustness of the negativity to the initial thermal occupation $\bar{N}$, which we assume to be in equilibrium with the environment for simplicity here but pre-cooling may be readily employed. While non-zero values of $\bar{N}$ reduce Wigner negativity, preparation of an initial mechanical ground state ($\bar{N}\rightarrow0$) is not a prerequisite for generating mechanical negativity.

\begin{figure}
    \centering
    \includegraphics[width=\columnwidth]{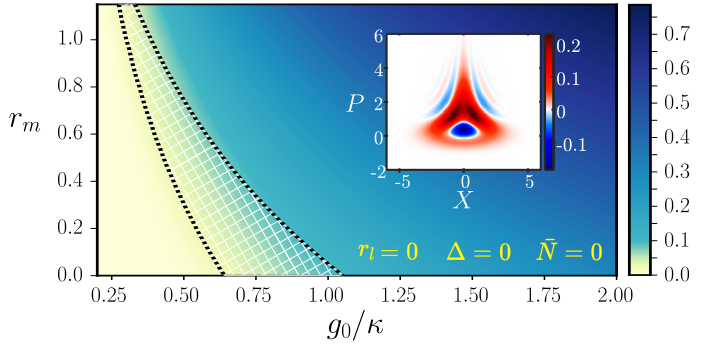}
    \caption{Photon counting following pulsed nonlinear cavity optomechanics increases mechanical negativity. Plot of the increase in the magnitude of the total negative volume due to single-photon detection compared to Fig.~\ref{fig_1_main}(e). Inset: mechanical Wigner function generated via single-photon detection with the same parameters as Fig.~\ref{fig_1_main}(b). Compared to Fig.~\ref{fig_1_main}(b), $\delta$ increases from $0.016$ to $0.39$. The cross-hatched region indicates where the negative volume is more than 1\% of the positive volume for the state generated via single-photon detection, \emph{but} less than 1\% of the positive volume for the mechanical state generated deterministically. 
    }
    \label{fig_2_main}
\end{figure}

%%%%%%%%%%%%%%%%%%%%%%%%%%%%%%%%%%%%%%%%%%%%%%%%%%%%%%%%%%%%%%%%%%%%%%%%%%%%%%%%%%%%%%%%%%%%%%%%%%%%%%%
\textit{Increasing mechanical negativity by photon counting.}---The Wigner negativity presented above is generated in a deterministic manner without utilizing measurement. Here, we explore how this negativity can be enhanced by combining this interaction with conditional photon-counting measurements on the optical mode.
If $n$ photons are detected after the nonlinear pulsed optomechanical interaction, the initial mechanical state $\rhoin$ transforms according to $\rho_{n}=\rme^{\rmi\varphi(X)n}\rhoin\rme^{-\rmi\varphi(X)n}$. Hence, the Wigner function of $\rho_{n}$ is $W_{n}(X,P)=\frac{1}{2\pi}\int_{-\infty}^{+\infty}\rme^{\rmi P u}\mathcal{K}_{n}(X,u)\bra{X-u/2}\rhoin\ket{X+u/2}\rmd{u}$ with $\mathcal{K}_{n}(X,u)=\left[f^{*}(X+u/2)f(X-u/2)\right]^n$. In the absence of optical loss, this transformation holds for any input optical state $\ket{\psi}$, where the probability to detect $n$ photons is $\Ph_{n}=|c_{n}|^2$, which depends on the optical input state.
In Fig.~\ref{fig_2_main}, the increase in $\delta$ due to single-photon detection is plotted as a function of $g_{0}/\kappa$ and $r_{m}$. 
Here, we see that, especially for small $r_{m}$, photon counting leads to a significant increase in mechanical negativity and is thus a valuable resource for generating nonclassicality. In the inset of Fig.~\ref{fig_2_main}, the mechanical Wigner function generated via single-photon detection is plotted using the same parameters as Fig.~\ref{fig_1_main}(b).
For these parameters, single-photon detection increases $\delta$ by a factor of $\sim20$ from $0.016$ to $0.39$. By comparing these two Wigner functions, we see that the Wigner function generated via photon counting occupies a smaller region of phase space and hence we expect this state to be more robust against mechanical decoherence. 
The nonclassical depth $\tau_{\mathrm{inf}}$~\cite{lee1991measure,lutkenhaus1995nonclassical} quantifies the number of thermal phonons needed to eliminate Wigner negativity, providing a direct measure of the experimental challenge in verifying Wigner negativity~\cite{milburn2016nonclassical}. In the Supplemental Material, we confirm that photon counting increases $\tau_{\mathrm{inf}}$, thus further providing a more experimentally accessible route to verify mechanical negativity. In contrast to the deterministic case, the mechanical states generated via photon counting are susceptible to optical loss and detection inefficiencies, with the resulting mechanical states depending on the optical input state--see the Supplementary Material for details~\cite{supp}.

%%%%%%%%%%%%%%%%%%%%%%%%%%%%%%%%%%%%%%%%%%%%%%%%%%%%%%%%%%
%%%%%%%%%%%%%%%%%%%%%%%% Figure 2 %%%%%%%%%%%%%%%%%%%%%%%%
%%%%%%%%%%%%%%%%%%%%%%%%%%%%%%%%%%%%%%%%%%%%%%%%%%%%%%%%%%
%%%%%%%%%%%%%%%%%%%%%%%%%%%%%%%%%%%%%%%%%%%%%%%%%%%%%%%%%%
%
%%%%%%%%%%%%%%%%%%%%%%%%%%%%%%%%%%%%%%%%%%%%%%%%%%%%%%%%%%

%%%%%%%%%%%%%%%%%%%%%%%%%%%%%%%%%%%%%%%%%%%%%%%%%%%%%%%%%%%%%%%%
\textit{Mechanical negativity via continuous driving.}---When the optomechanical system is driven continuously by a coherent optical input and no measurement is made on the optical field, the master equation for the mechanical mode is
\begin{eqnarray}
    \dfrac{\rmd{\rho}}{\rmd{t}}&=&-\frac{\rmi}{\hbar}[H_{0},\rho]+2\gamma\left(\bar{N}+1\right)\mathcal{D}\left[b\right]\rho\nonumber\\&+&2\gamma\bar{N}\mathcal{D}\left[b^{\dagger}\right]\rho+2k\mathcal{D}\left[f(X)\right]\rho.\label{eqmain:ME}
\end{eqnarray}
Here, $H_{0}=\hbar\wm b^{\dagger}b$ is the free mechanical Hamiltonian, $\gamma$ is the mechanical decay rate, $\bar{N}$ is the mean occupation of the thermal environment, $\mathcal{D}$ is the Lindblad superoperator, and $2k$ is the constant input photon flux. 
In this section we are interested in steady-state solutions to Eq.~\eqref{eqmain:ME} due to their inherent stability and experimental simplicity.
%  

%%%%%%%%%%%%%%%%%%%%%%%%%%%%%%%%%%%%%%%%%%%%%%%%%%%%%%%%%%%%%%%%
%%%%%%%%%%%%%%%% Figure 3 %%%%%%%%%%%%%%%%%%%%%%%%%%%%%%%%%%%%%%
%%%%%%%%%%%%%%%%%%%%%%%%%%%%%%%%%%%%%%%%%%%%%%%%%%%%%%%%%%%%%%%%

\begin{figure*}
    \centering
    \includegraphics[width=\textwidth]{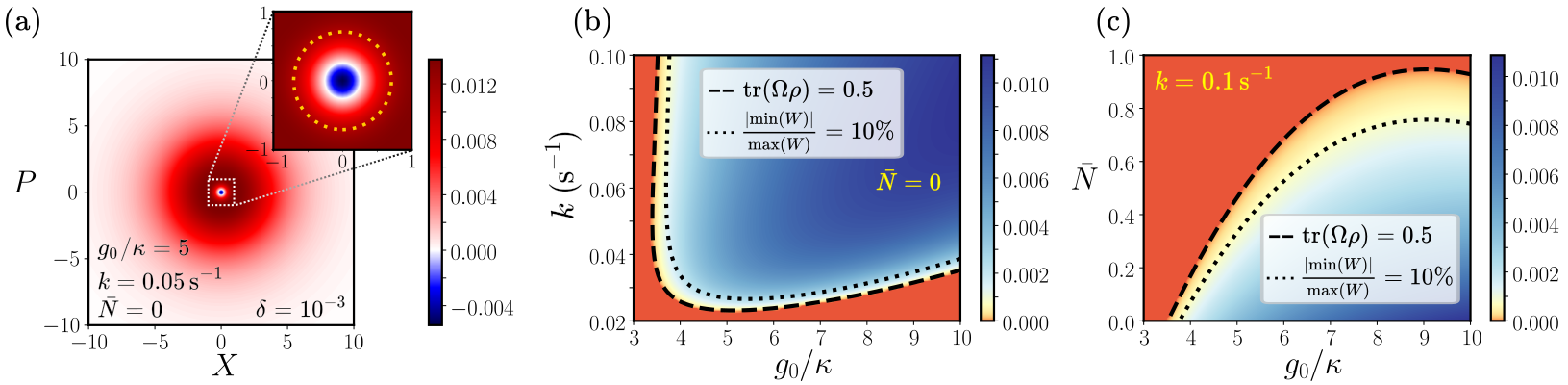}
    \caption{Steady-state mechanical Wigner negativity via continuously-driven nonlinear cavity optomechanics. (a) Mechanical Wigner function for a state with $g_{0}/\kappa=5$, photon flux parameter $k=0.05\,\mathrm{s}^{-1}$, and $\bar{N}=0$. Inset: zoom-in of the Wigner function near the origin where the radius of the dashed circle is the ground-state width. (b) The absolute value of the minimum of the mechanical Wigner function $|\mathrm{min}(W)|$ plotted as a function of the photon flux $k$ and the ratio $g_{0}/\kappa$. The dashed curve indicates the boundary where the negativity witness $\mathrm{tr}(\Omega\rho)$ exceeds the threshold value of $0.5$ and the dotted curve indicates the boundary beyond which $|\mathrm{min}(W)|$ exceeds $10\%$ of the maximum value of the Wigner function. (c) $|\mathrm{min}(W)|$ plotted as a function of the occupation of the mechanical thermal environment $\bar{N}$ and $g_{0}/\kappa$. In these plots, the mechanical damping rate is $\gamma/2\pi=10^{-3}\,\mathrm{Hz}$ and a mechanical frequency of $\wm/2\pi=10^{5}\,\mathrm{Hz}$ is chosen to ensure the rotating wave approximation is valid within the parameter range explored.
    }
    \label{fig_5_main}
\end{figure*}
%%%%%%%%%%%%%%%%%%%%%%%%%%%%%%%%%%%%%%%%%%%%%%%%%%%%%%%%%%%%%%%%
%%%%%%%%%%%%%%%%%%%%%%%%%%%%%%%%%%%%%%%%%%%%%%%%%%%%%%%%%%%%%%%%
Fig.~\ref{fig_5_main}(a) plots the Wigner function of the mechanical steady state under continuous drive in this nonlinear regime. This state solves Eq.~\eqref{eqmain:ME} for $\dot{\rho}=0$ and is valid when the mechanical frequency dominates over all other relevant rates in Eq.~\eqref{eqmain:ME}.
The resulting states are rotationally invariant in phase space and exhibit negativity resembling a \emph{pinhole} centred at the origin of phase space.
To verify this Wigner negativity without using full quantum state tomography, we consider the witness $\Omega=\sum_{k=0}^{n-1}\ket{2k+1}\bra{2k+1}$, with $n\in \mathbb{N}$~\cite{chabaud2021certification,chabaud2021witnessing}, which may be measured with phonon-counting techniques~\cite{vanner2015towards}. The quantity $\mathrm{tr}(\Omega\rho)$ gives the sum of the fidelities with the odd Fock states, which are all minimum at the origin of phase space, and if $\mathrm{tr}(\Omega\rho)>0.5$ the state possesses Wigner negativity. 
Due to this connection between the witness and the minimum of the Wigner function, in Fig.~\ref{fig_5_main}(b\&c) we plot the absolute value of the minimum of the Wigner function $|\mathrm{min}(W)|$. The negative volume indicator is also plotted in the Supplemental Material~\cite{supp}, which shows qualitatively similar behaviour.
In Fig.~\ref{fig_5_main}(b), $|\mathrm{min}(W)|$ is plotted as a function of $g_{0}/\kappa$ and the photon flux parameter $k$ for a thermal environment at $\bar{N}\rightarrow0$. Here, $\mathrm{tr}(\Omega\rho)>0.5$ for $g_{0}/\kappa>3.4$, simulated for $n=100$, indicating that single-photon strong coupling is required to observe mechanical Wigner negativity in the steady state. However, the negativity at the origin quickly grows with $g_{0}/\kappa$ and the dotted curve indicates when the minimum of the Wigner function begins to exceed $10\%$ of its maximum value. Fig.~\ref{fig_5_main}(c) investigates the deleterious effect of the thermal occupation of the environment on mechanical negativity. For the parameters explored here, we see that $\mathrm{tr}(\Omega\rho)>0.5$ is possible for a bath thermal occupation of $\bar{N}\lesssim 1$. Note that, as Eq.~\eqref{eqmain:ME} is Markovian, the steady state is independent of the mechanical initial conditions.

%%%%%%%%%%%%%%%%%%%%%%%%%%%%%%%%%%%%%%%%%%%%%%%%%%%%%%%%%%%%%%%%%
\textit{Conclusions and Outlook.}---We demonstrate that, counterintuitively, mechanical Wigner negativity may be generated deterministically in the unresolved-sideband regime of cavity quantum optomechanics without any additional nonlinearities beyond radiation-pressure inside an optical cavity. This nonlinear operation arises from the nonlinearity of the cavity response with mechanical position itself, which we utilize to generate non-Gaussian mechanical quantum states.
By driving the optomechanical cavity with short optical pulses, and employing optical and mechanical squeezing operations, mechanical negativity may be generated even when single-photon strong coupling and ground-state cooling requirements are relaxed. For continuous optical drives, steady-state negativity persists in the single-photon strong coupling regime and in the presence of finite thermal occupation. 

The nonclassical state generation proposed here can be implemented in a wide array of present-day and near-future experiments operating in the unresolved-sideband regime. For instance, ultracold-atom-based optomechanical systems now operate in both the unresolved-sideband regime and the single-photon strong coupling regime~\cite{brennecke2008cavity,purdy2010tunable} and provide a promising platform to readily explore the physics proposed here. Moreover, improvements to the sliced photonic crystal devices reported in Refs~\cite{leijssen2017nonlinear, muhonen2019state}, which have achieved $g_{0}/\kappa \approx 3\times10^{-3}$, also provide a promising route to observe the mechanical negativity via pulsed optomechanics proposed here given the large values of $g_0$ and the prospect to reduce $\kappa$ in these systems.

This work deepens our understanding of nonlinear cavity quantum optomechanics and sheds light on the important role of the cavity above and below the single-photon strong-coupling regime. By harnessing these nonlinearites, our work opens new experimental avenues for quantum operations and deterministic quantum state engineering of mechanical degrees of freedom.
Furthermore, this work provides a foundation for a wide range of further studies including enhancing the steady-state negativity generated here via mechanical squeezing achieved via parametric amplification~\cite{rugar1991mechanical}, and mechanical multi-mode extensions to this model to explore the role of thermal intermodulation noise~\cite{fedorov2020thermal,pluchar2023thermal} in a quantum regime.

\vspace{1mm}
\begin{acknowledgments}
\textit{Acknowledgements.}---We acknowledge useful discussions with Amaya Calvo-Sanchez, Daniel Carney, Rufus Clarke, Evan Cryer-Jenkins, Radim Filip, Arjun Gupta, Kiran E. Khosla, Florian Marquardt, and Gerard J. Milburn. This project was supported by UK Research and Innovation (MR/S032924/1, MR/X024105/1), the Engineering and Physical Sciences Research Council (EP/T031271/1, and a Doctoral Prize Fellowship awarded to J.C.), the Science and Technology Facilities Council (ST/W006553/1), the Netherlands Organisation for Scientific Research (NWO), and the European Research Council (ERC) Starting Grant (no. 759644-TOPP) and Consolidator Grant (no. 101088055-Q-MEME).
\end{acknowledgments}

% % with bibfile
% \bibliography{bib1}

% without bibfile
\providecommand{\noopsort}[1]{}\providecommand{\singleletter}[1]{#1}%
%

%%%%%%%%%%%%%%%%%%%%%%%%%%%%%%%%%%%%%%%%%%%%%%%%%%%%%%%%%%%%%%%%%%%%%%%%%%%%%%%%%%%%%%%%%%%%%%%%%%%%%%%%%%%%%%%%%%%%%%%%%%%%%%%%%%%%%%%%%%%%%%%%%%%%%%%%%%%%%%%%

%%%%%%%%%%%%%%%%%%%%%%%%%%%%%%%%%%%%%%%%%%%%%%%%%%%%%%%%%%%%%%%%%
%%%%% SUPPLEMENTAL %%%%%
%%%%%%%%%%%%%%%%%%%%%%%%%%%%%%%%%%%%%%%%%%%%%%%%%%%%%%%%%%%%%%%%%

\onecolumngrid
\clearpage
\newgeometry{left=1.5cm,right=1.5cm,top=1.5cm,bottom=1.5cm}
\setcounter{equation}{0}
\def\theequation{S\arabic{equation}}
\setcounter{figure}{0}
\renewcommand{\thefigure}{S\arabic{figure}}
\pagenumbering{roman}

\begin{center}
\textbf{\large{Deterministic Mechanical Wigner Negativity via Nonlinear Cavity Quantum Optomechanics in the Unresolved-Sideband Regime:\\ Supplemental Material}}
\end{center}
\vspace{10pt}

\begin{centering}
Jack~Clarke\,\orcidlink{0000-0001-8055-449X},$^{1, \, 2}$~  
Pascal Neveu\,\orcidlink{0000-0002-7671-2763},$^{3}$~
Ewold Verhagen\,\orcidlink{0000-0002-0276-8430},$^{3}$~
and Michael R. Vanner\,\orcidlink{0000-0001-9816-5994}$^{1}$\\
\end{centering}

\vspace{16pt}

\begin{centering}
\textit{\small
 $^1$Quantum Measurement Lab, Blackett Laboratory, Imperial College London, London SW7 2BW, United Kingdom\\
$^2$ Department of Physics and Astronomy, University College London, Gower Street, WC1E 6BT London, United Kingdom \\
 $^3$Center for Nanophotonics, AMOLF, Science Park 104, 1098 XG Amsterdam, The Netherlands\\
}
\end{centering}

\begin{quote}
{\small{In this Supplemental, we provide further details on the generation of mechanical Wigner negativity via nonlinear cavity quantum optomechanics and derive the equations presented in the main text. Firstly,  we review the nonlinear optical response of the cavity and derive the expression for the mechanical states created via nonlinear pulsed optomechanics. We then give expressions for the modulus-squared of the coefficients $c_{n}$ when the optical input pulse is either in the coherent state or a the squeezed vacuum state. Following this, we derive the Wigner function for the mechanical state created in absence of the nonlinear cavity response. After this, we describe how photon counting measurements can enhance mechanical Wigner negativity, before making a connection to Hudson's theorem. Finally, we derive the deterministic master equation corresponding to a completely general continuous Gaussian optical drive, which allows one to find the mechanical steady state.}}
\end{quote}

\vspace{16pt}
\twocolumngrid
\noindent\textbf{Contents}\\
{\small{
\noindent I. \hyperlink{S1}{Nonlinear optical response}\dotfill{\pageref{sec:S1}} \\
\noindent II. \hyperlink{S2}{Pulsed optomechanics}\dotfill{\pageref{sec:S2}} \\
\hspace*{10mm} A. \hyperlink{S2A}{Mechanical Wigner function} \dotfill{\pageref{sec:S2A}} \\
\hspace*{10mm} B. \hyperlink{S2B}{The effect of optical detuning} \dotfill{\pageref{sec:S2B}} \\
\hspace*{10mm} C. \hyperlink{S2C}{Without the full cavity response} \dotfill{\pageref{sec:S2C}} \\
\noindent III. \hyperlink{S3}{Increasing mechanical negativity by photon counting}\dotfill{\pageref{sec:S3}} \\
\hspace*{10mm} A. \hyperlink{S3A}{Mechanical Wigner function} \dotfill{\pageref{sec:S3A}} \\
\hspace*{10mm} B. \hyperlink{S3B}{Connection to Hudson's theorem} \dotfill{\pageref{sec:S3B}} \\
\hspace*{10mm} C. \hyperlink{S3C}{Single-photon detection} \dotfill{\pageref{sec:S3C}} \\
\noindent IV. \hyperlink{S4}{Environmental effects}\dotfill{\pageref{sec:S4}} \\
\hspace*{10mm} A. \hyperlink{S4A}{Optical loss} \dotfill{\pageref{sec:S4A}} \\
\hspace*{10mm} B. \hyperlink{S4B}{Mechanical thermal effects \& nonclassical depth} \dotfill{\pageref{sec:S4B}} \\
\noindent V. \hyperlink{S5}{Continuously-driven optomechanics}\dotfill{\pageref{sec:S5}} \\
\hspace*{10mm} A. \hyperlink{S5A}{General Gaussian input optical state} \dotfill{\pageref{sec:S5A}} \\
\hspace*{10mm} B. \hyperlink{S5B}{The master equation} \dotfill{\pageref{sec:S5B}} \\
\hspace*{10mm} C. \hyperlink{S5C}{Steady state \& the rotating wave approximation} \dotfill{\pageref{sec:S5C}} \\
}}

\onecolumngrid

%%%%%%%%%%%%%%%%%%%%%%%%%%%%%%%%%%%%%%%%%%%%%%%%%%%%%%%%%%%%%%%%%
\hypertarget{S1}{}\section{I. Nonlinear optical response}\label{sec:S1}
For completeness, we first give a discussion of the nonlinear optical response so that one can see a derivation of $f(X)$ here. In a frame rotating at the cavity frequency $\wc$, the Heisenberg-Langevin equations describing pulsed optomechanics in the unresolved sideband regime ($\kappa\gg\wm$) are 
\begin{eqnarray}
\dot{a}&=&\rmi(\sqrt{2}g_{0}X+\Delta)a-\kappa{a}+\sqrt{2\kappa}\ain\label{HLcav}\\
\dot{X}&\simeq&0\\
\dot{P}&\simeq&\sqrt{2}g_{0}a^{\dagger}a\label{HLmomentum}.
\end{eqnarray}
Here, the unnormalized detuning is $\Delta=\wl-\wc$, where $\wl$ is the optical drive carrier angular frequency, and free mechanical evolution and dissipation may be neglected as the interaction time is much less than the mechanical period. The cavity mode may be adiabatically eliminated ($\dot{a}\simeq0$) from the dynamics when operating in the unresolved sideband regime and when the characteristic timescale $\tau$ on which the optical input varies satisfies $\kappa\gg\tau^{-1}$. In this case, the input-output relation for the optical cavity $\aout=\ain-\sqrt{2\kappa}\ain$ becomes $\aout=f(X)\,\ain$, where $f(X)$ is the nonlinear optical response function 
\begin{equation}
    f(X)=\dfrac{1+\rmi\left(\frac{\mu}{2}{X}+\Db\right)}{1-\rmi\left(\frac{\mu}{2}{X}+\Db\right)},
\end{equation}
and $\Db=\Delta/\kappa$. The input-output relation can be written explicitly as a unitary transformation, $\aout=U^{\dagger}\ain U$, where $U=\rme^{\rmi\varphi(X)\nl}$ and $\varphi(X)=\arg(f)$. This nonlinear response function may be used to describe how the amplitude and phase of the light are transformed by the interaction, and also how the mechanical quadratures vary through use of the unitary $U$.

%%%%%%%%%%%%%%%%%%%%%%%%%%%%%%%%%%%%%%%%%%%%%%%%%%%%%%%%%%%%%%%%%
\hypertarget{S2}{}\section{II. Pulsed optomechanics}\label{sec:S2}

%%%%%%%%%%%%%%%%%%%%%%%%%%%%%%%%%%%%%%%%%%%%%%%%%%%%%%%%%%%%%%%%%
\hypertarget{S2A}{}\subsection{A. Mechanical Wigner function}\label{sec:S2A}
The mechanical state generated deterministically via the nonlinear cavity optomechanical interaction between a general pure optical input state $\ket{\psi}=\sum_{n=0}^{\infty}c_{n}\ket{n}$ and the initial mechanical Gaussian state $\rhoin$ is
\begin{eqnarray}
\rhof&=&\tr_{l}\left(U\rhoin\otimes\ket{\psi}\bra{\psi} U^{\dagger}\right)\\
&=&\sum_{n=0}^{\infty}|c_{n}|^2\rme^{\rmi\varphi(X)n}\rhoin\rme^{-\rmi\varphi(X)n}.\label{seq:rho_f}
\end{eqnarray}
The Wigner function of the state $\rhof$ is then 
 \begin{eqnarray}
  W(X,P)&=&\frac{1}{2\pi}\int_{-\infty}^{+\infty}~\rme^{\rmi Pu}\bra{X-u/2}\rhof\ket{X+u/2}~\rmd{u}\nonumber\\
  &=&\frac{1}{2\pi}\int_{-\infty}^{+\infty}~\rme^{\rmi Pu}~\mathcal{K}(X,u)\bra{X-u/2}\rhoin\ket{X+u/2}~\rmd{u},
 \end{eqnarray}
 where the kernel function $ \mathcal{K}(X,u)$ is
 \begin{eqnarray}
 \mathcal{K}(X,u)&=&\sum_{n=0}^{\infty}|c_{n}|^2\rme^{\rmi\varphi(X-u/2)n}\rme^{-\rmi\varphi(X+u/2)n}\\
&=&\sum_{n=0}^{\infty}|c_{n}|^2\left[f(X-u/2)f^{*}(X+u/2)\right]^n.
 \end{eqnarray}
 
 It is useful to note that for an initial mechanical Gaussian state with first moments $(X_{G},P_{G})^{\mathrm{T}}$, position variance $V_{X}$, momentum variance $V_{P}$, position-momentum covariance $V_{XP}$, and covariance-matrix determinant $d=V_{X}V_{P}-V_{XP}^2$, one has that
 \begin{eqnarray}
 \bra{X-u/2}\rhoin\ket{X+u/2}&=&\dfrac{1}{\sqrt{2\pi V_{X}}}\exp\left[-\frac{d}{2V_{X}}u^2-\frac{1}{2V_{X}}(X-X_{G})^2-\rmi P_{G}u-\rmi\frac{V_{XP}}{V_{X}}(X-X_{G})u\right].
 \end{eqnarray}
For an initial thermal mechanical state with occupation $\bar{n}$, mechanical momentum squeezing initializes the state to $(X_{G},P_{G})^{\mathrm{T}}=(0,0)^{\mathrm{T}}$, $V_{X}=(\bar{n}+1/2)\rme^{+2r_{m}}$, $V_{P}=(\bar{n}+1/2)\rme^{-2r_{m}}$, and $V_{XP}=0$.

\subsubsection{Coherent optical input state}
When the input optical state is a coherent state $\ket{\psi}=\ket{\alpha}$, with $\ket{\alpha}=D(\alpha)$ and $D(\alpha)=\rme^{\alpha a^{\dagger}-\alpha^{*}a}$,  the coefficients are $c_{n}=\rme^{-|\alpha|^2/2}\alpha/\sqrt{n!}$ and so the mechanical state after the optomechanical interaction is 
\begin{eqnarray}
\rhof&=&\rme^{-|\alpha|^2}\sum_{n=0}^{\infty}\dfrac{|\alpha|^{2n}}{n!}\rme^{\rmi\varphi(X)n}\rhoin\rme^{-\rmi\varphi(X)n},\label{eqsupp:rhof_alpha}
\end{eqnarray}
 and the Wigner function is
 \begin{eqnarray}
  W(X,P)  &=&\frac{1}{2\pi}\int_{-\infty}^{+\infty}~\rme^{\rmi (P-\mathcal{S})u}\bra{X-u/2}\rhoin\ket{X+u/2}~\rmd{u}\label{eqsupp:Wf_alpha},
 \end{eqnarray}
 where we have written $\mathcal{K}(X,u)=\rme^{-\rmi\mathcal{S}u}$. Inserting Eq.~\eqref{eqsupp:rhof_alpha} into Eq.~\eqref{eqsupp:Wf_alpha} allows one to identify that
 \begin{eqnarray}
 \rme^{-\rmi\mathcal{S}u}&=&\rme^{-|\alpha|^2}\sum_{n=0}^{\infty}\dfrac{|\alpha|^{2n}}{n!}\rme^{\rmi\varphi(X-u/2)n}\rme^{-\rmi\varphi(X+u/2)n}\nonumber\\
&=&\rme^{-|\alpha|^2}\sum_{n=0}^{\infty}\dfrac{|\alpha|^{2n}}{n!}\left[f^{*}(X+u/2)f(X-u/2)\right]^n\nonumber\\
&=&\exp\left\{-|\alpha|^2\left[1-f^*(X+u/2)f(X-u/2)\right]\right\}\nonumber\\
&=&\exp\left\{-|\alpha|^2\dfrac{\rmi\mu u}{\left[\frac{\mu}{2}(X+\frac{u}{2})+\Db-\rmi\right]\left[\frac{\mu}{2}(X-\frac{u}{2})+\Db+\rmi\right]}\right\},
 \end{eqnarray}
and hence
\begin{eqnarray}
\mathcal{S}
&=&|\alpha|^2\dfrac{\mu}{\left[\frac{\mu}{2}(X+\frac{u}{2})+\Db-\rmi\right]\left[\frac{\mu}{2}(X-\frac{u}{2})+\Db+\rmi\right]}.
\end{eqnarray}

\subsubsection{Squeezed vacuum optical input state}
When the input optical state is a squeezed vacuum $\ket{\psi}=S(\zeta)\ket{0}$, with $S(\zeta)=\rme^{\frac{1}{2}(\zeta^{*} a^2-\zeta a^{\dagger 2})}$ and $\zeta=r_{l}\rme^{\rmi\theta}$, the coefficients in the number basis are given by
\begin{eqnarray}
c_{2n+1}&=&0,\\
c_{2n}&=&\dfrac{1}{\sqrt{\cosh r_{l}}}\left(-\frac{1}{2}\rme^{\rmi\theta}\tanh r_{l} \right)^n\dfrac{\sqrt{(2n)!}}{n!}.
\end{eqnarray}
In this case, the mechanical state after the optomechanical interaction is
\begin{eqnarray}\label{eq:mechanical_state_sq_vac_input}
\rhof&=&\dfrac{1}{\cosh r_{l}}\sum_{n=0}^{\infty}\left(\frac{1}{2}\tanh r_{l}\right)^{2n}\dfrac{(2n)!}{(n!)^2}~\rme^{\rmi\varphi(X)n}\rhoin\rme^{-\rmi\varphi(X)n}
\end{eqnarray}
and the kernel inside the integral for the Wigner function is given by
\begin{eqnarray}
 \mathcal{K}(X,u)&=&\dfrac{1}{\cosh r_{l}}\sum_{n=0}^{\infty}\left(\frac{1}{2}\tanh r_{l}\right)^{2n}\dfrac{(2n)!}{(n!)^2}\left[f(X-u/2)f^{*}(X+u/2)\right]^{2n}\label{eq:kernel_squeezed_vac}.
\end{eqnarray}
Note that Eqs~\eqref{eq:mechanical_state_sq_vac_input} and \eqref{eq:kernel_squeezed_vac} are independent of the squeezing angle $\theta$. Thus, unlike the squeezing operation applied to the initial mechanical state, the direction of the optical squeezing does not affect the final mechanical state. Using the relation $(2n)!=2^{n}n!(2n-1)!!$ and the sum
\begin{eqnarray}
\sum_{n=0}^{\infty}\dfrac{(2n-1)!!}{n!}\left(\frac{x}{2}\right)^n=\dfrac{1}{\sqrt{1-x}},
\end{eqnarray}
Eq.~\eqref{eq:kernel_squeezed_vac} can be rewritten as
\begin{eqnarray}
\mathcal{K}(X,u)&=&\dfrac{1}{\cosh r_{l}}\dfrac{1}{\sqrt{1-\tanh^2r_{l}\left[f(X-u/2)f^{*}(X+u/2)\right]^{2}}}.
\end{eqnarray}

%%%%%%%%%%%%%%%%%%%%%%%%%%%%%%%%%%%%%%%%%%%%%%%%%%%%%%%%%%%%%%%%%
%%%%%%%%%%%%%%%%%%%%%%%%%%%% Figure detuning %%%%%%%%%%%%%%%%%%%%
%%%%%%%%%%%%%%%%%%%%%%%%%%%%%%%%%%%%%%%%%%%%%%%%%%%%%%%%%%%%%%%%%
\begin{figure*}
    \centering
    \includegraphics[width=\textwidth]{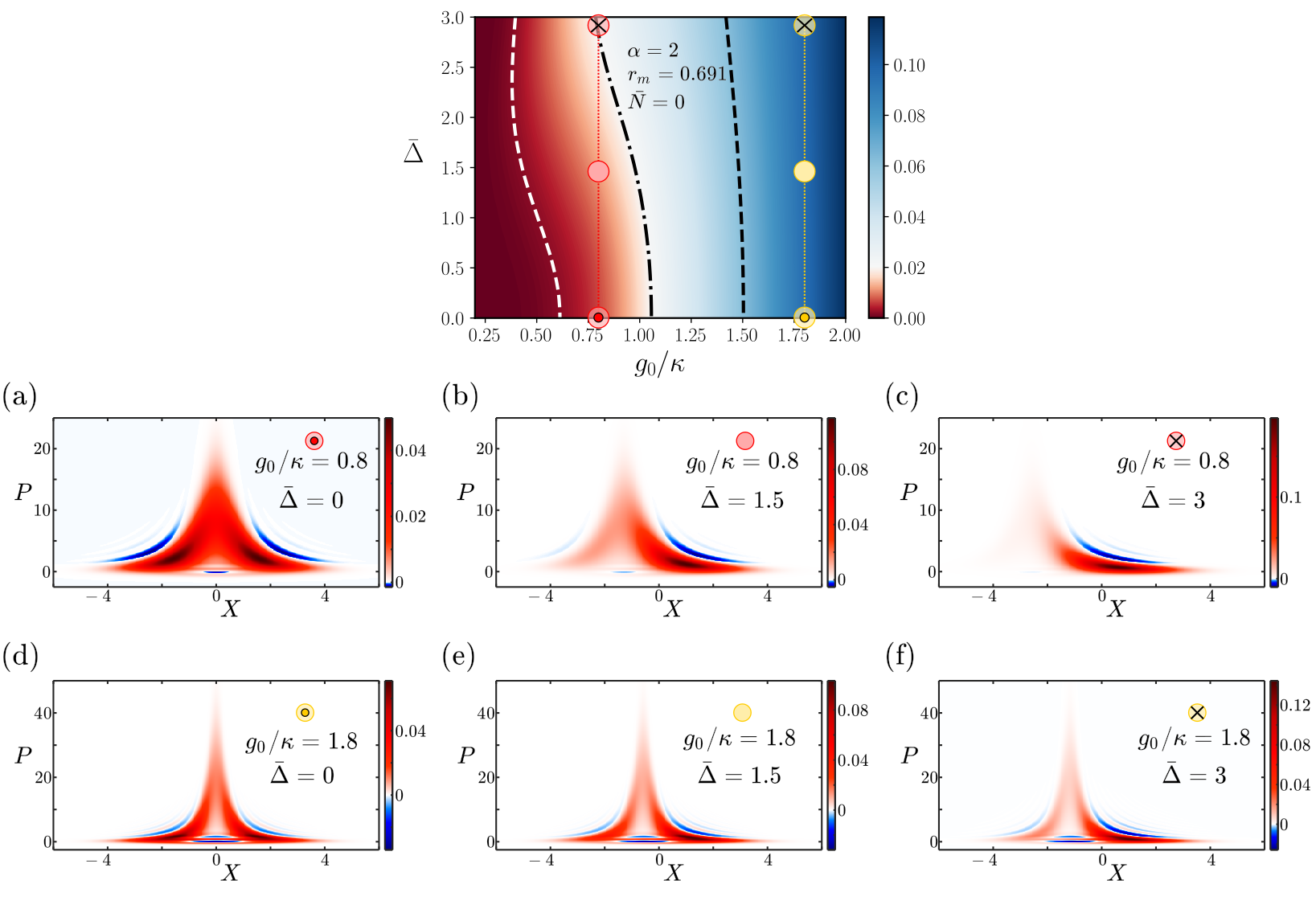}
    \caption{Comparison between the mechanical states generated with $g_{0}/\kappa=0.8$ versus $g_{0}/\kappa=1.8$ to demonstrate the effect of the optical detuning $\bar{\Delta}$. Top: Fig.~\ref{fig_1_main}(d) from the main text 
    with two vertical lines superimposed on the plot at $g_{0}/\kappa=0.8$ (red) and $g_{0}/\kappa=1.8$ (yellow). (a) $g_{0}/\kappa=0.8$, $\bar{\Delta}=0$: $\delta=0.00648$ and $\mathrm{min}(W)=-0.00141 $ located at $X=\pm2.4533$ and $P=3.1617$. Due to the symmetry of the state about $X=0$, there are two minima.
    (b) $g_{0}/\kappa=0.8$, $\bar{\Delta}=1.5$: $\delta=0.01190$ and $\mathrm{min}(W)=-0.00520 $ located at $X=1.4933$ and $P=2.6162$. The detuning now introduces a left-right asymmetry and biases the negativity towards the right.
    (c) $g_{0}/\kappa=0.8$, $\bar{\Delta}=3$: $\delta=0.02091$ and $\mathrm{min}(W)=-0.01013 $ located at $X=0.5333$ and $P=2.2196$. The negativity is further enhanced and concentrated towards the right. 
    (d) $g_{0}/\kappa=1.8$, $\bar{\Delta}=0$: $\delta=0.09507$ and $\mathrm{min}(W)=-0.03401 $ located at $X=0$ and $P=0.1371$. At higher values of $g_{0}/\kappa$, the negativity is concentrated more around the central region. 
    (e) $g_{0}/\kappa=1.8$, $\bar{\Delta}=1.5$: $\delta=0.09602$ and $\mathrm{min}(W)=-0.03171 $ located at $X=-0.4800$ and $P=0.1371$. While the detuning shifts the central negative region slightly to the left, the Wigner negativity is still concentrated in this central region. (f) $g_{0}/\kappa=1.8$, $\bar{\Delta}=3$: $\delta=0.09790$ and $\mathrm{min}(W)=-0.02569$ located at $X=-0.9600$ and $P=0.1371$. In this high $g_{0}/\kappa$ regime, the value of $\delta$ increases only marginally with the detuning $\bar{\Delta}$, while the value of $\mathrm{min}(W)$ actually decreases.}
    \label{appendix_fig_1}
\end{figure*}
%%%%%%%%%%%%%%%%%%%%%%%%%%%%%%%%%%%%%%%%%%%%%%%%%%%%%%%%%%%%%%%%%
%
%%%%%%%%%%%%%%%%%%%%%%%%%%%%%%%%%%%%%%%%%%%%%%%%%%%%%%%%%%%%%%%%%

%%%%%%%%%%%%%%%%%%%%%%%%%%%%%%%%%%%%%%%%%%%%%%%%%%%%%%%%%%%%%%%%%
\hypertarget{S2B}{}\subsection{B. The effect of optical detuning}\label{sec:S2B}

In Fig.~\ref{appendix_fig_1}, we study the behaviour of the plot in Fig.~\ref{fig_1_main}(d) in more detail. Specifically, we study why the negative volume $\delta$ depends more strongly on the detuning $\bar{\Delta}$ for lower values of $g_{0}/\kappa$ than for higher values. In summary, at lower values of $g_{0}/\kappa$, negativity is concentrated more strongly around the left and right negative lobes of the mechanical Wigner function. Meanwhile at higher values of $g_{0}/\kappa$, the Wigner negativity is distributed more around the central negative region. This difference in the distribution of Wigner negativity may be seen by comparing Figs~\ref{appendix_fig_1}(a) and \ref{appendix_fig_1}(d), for instance. Selecting a non-zero detuning value $\bar{\Delta}$ introduces a left-right asymmetry along the $X$-axis of the Wigner function. For the states with lower values of $g_{0}/\kappa$, this asymmetry may be used to bias the negativity from being equally distributed between left and right negative lobes to being more concentrated towards one side. Therefore, the effect of the detuning is to enhance and concentrate the Wigner negativity, which greatly increases both $\delta$ and $\mathrm{min}(W)$---see Figs~\ref{appendix_fig_1}(a), \ref{appendix_fig_1}(b), and \ref{appendix_fig_1}(c). A non-zero detuning induces the same asymmetry in states with higher $g_{0}/\kappa$, however, as the Wigner negativity is mostly concentrated around the central negative region, biasing Wigner negativity towards one side has little effect on the value of $\delta$---see Figs~\ref{appendix_fig_1}(d), \ref{appendix_fig_1}(e), and \ref{appendix_fig_1}(f).

%%%%%%%%%%%%%%%%%%%%%%%%%%%%%%%%%%%%%%%%%%%%%%%%%%%%%%%%%%%%%%%%%
\hypertarget{S2C}{}\subsection{C. Without the full cavity response}\label{sec:S2C}
If the nonlinearity of radiation pressure $\left[H/\hbar=-{g}_{0}{a}^{\dagger}a(b+b^{\dagger})\right]$ is accounted for, but the nonlinear response of the cavity is ignored, the optomechanical interaction is described by $\rme^{\rmi(\mu X+2\Db)\nl}$. [Note to first order in $X$, $\varphi(X)\approx(\mu X+2\Db)$.]
In this case, if no measurement is made on the output optical state, the final mechanical state is given by 
\begin{eqnarray}
\rhof'&=&\tr_{l}\left(\rme^{\rmi(\mu X+2\Db)\nl}\rhoin \rme^{-\rmi(\mu X+2\Db)\nl}\right),\nonumber\\
&=&\sum_{n=0}^{\infty}|c_{n}|^2\rme^{\rmi n\mu X}\rhoin\rme^{-\rmi n\mu X},
\end{eqnarray}
instead of Eq.~\eqref{seq:rho_f}. The Wigner function of $\rhof'$ is given by
\begin{eqnarray}
W'&=&\frac{1}{2\pi}\sum_{n=0}^{\infty}|c_{n}|^2\int_{-\infty}^{+\infty}~\rme^{\rmi (P-n\mu)u}\bra{X-u/2}\rhoin\ket{X+u/2}~\rmd{u}\nonumber\\
&=&\sum_{n=0}^{\infty}|c_{n}|^2 W_{i}(X,P-n\mu),
\end{eqnarray}
which demonstrates that $W'$ is a statistical mixture of displaced copies of the initial Gaussian distribution $W_{i}$ and thus cannot possess any Wigner negativity.

%%%%%%%%%%%%%%%%%%%%%%%%%%%%%%%%%%%%%%%%%%%%%%%%%%%%%%%%%%%%%%%%%
\hypertarget{S3}{}\section{III. Increasing mechanical negativity by photon counting}\label{sec:S3}

%%%%%%%%%%%%%%%%%%%%%%%%%%%%%%%%%%%%%%%%%%%%%%%%%%%%%%%%%%%%%%%%%
\hypertarget{S3A}{}\subsection{A. Mechanical Wigner function}\label{sec:S3A}
For the general optical input state $\ket{\psi}=\sum_{n=0}^{\infty}c_{n}\ket{n}$, the mechanical state generated after the nonlinear cavity optomechanical interaction followed by the detection of $n$ photons is
\begin{eqnarray}
\rho_{n}=\dfrac{\Upsilon_{n}\rhoin\Upsilon_{n}^{\dagger}}{\Ph_{n}},
\end{eqnarray}
where $\Upsilon_{n}=\bra{n}U\ket{\psi}=c_{n}\,\rme^{\rmi \varphi(X) n}$ and $\Ph_{n}=\tr\left(\Upsilon_{n}^{\dagger}\Upsilon_{n}\rhoin\right)=|c_{n}|^2$. Hence, the mechanical state after detecting $n$ photons can be written as
\begin{eqnarray}
\rho_{n}&=&\rme^{\rmi\varphi(X)n}\rhoin\rme^{-\rmi\varphi(X)n}.
\end{eqnarray}
This illustrates how the state of the mechanical mode does not depend on the choice of the input optical state $\ket{\psi}$, however, the heralding probability $\Ph_{n}$ does depend on $\ket{\psi}$.

When the optical input state is equal to the coherent state, $\ket{\psi}=\ket{\alpha}$, the heralding probability is $\Ph_{n}=\rme^{-|\alpha|^{2}}|\alpha|^{2n}/n!$. In this case, the maximum value of $\Ph_{n}$ is found at $|\alpha|=\sqrt{n}$ and is given by $\mathrm{max}(\Ph_{n})=\rme^{-n}n^{n}/n!$. Likewise, when the optical input state is the squeezed vacuum state, $\ket{\psi}=S(\zeta)\ket{0}$, the heralding probability is given by
\begin{eqnarray}
\Ph_{n}=\begin{cases}
  \dfrac{1}{\cosh r}\left(\frac{1}{2}\tanh r\right)^{2n}\dfrac{(2n)!}{(n!)^2}  & n \text{ is even} \\
  0 & n \text{ is odd}.
\end{cases}
\end{eqnarray}
For even $n$, the heralding probability is maximized at $r=\mathrm{arcsinh}(\sqrt{2n})$, which leads to $\mathrm{max}(\Ph_{2n})=\frac{1}{\sqrt{1+2n}}\left(\frac{1}{2}\sqrt{\frac{2n}{1+2n}}\right)^{2n}\frac{(2n)!}{(n!)^2}$.

Regardless of the choice of $\ket{\psi}$, the Wigner function of $\rho_{n}$ is
\begin{eqnarray}
W_{n}(X,P)&=&\frac{1}{2\pi}\int_{-\infty}^{+\infty}~\rme^{\rmi Pu}\bra{X-u/2}\rho_{n}\ket{X+u/2}~\rmd{u}\nonumber\\
&=&\frac{1}{2\pi}\int_{-\infty}^{+\infty}~\rme^{\rmi Pu}\mathcal{K}_{n}(X,u)\bra{X-u/2}\rhoin\ket{X+u/2},~\rmd{u}\label{eqsupp:W_n}
\end{eqnarray}
with $\mathcal{K}_{n}(X,u)=\left[f^{*}(X+u/2)f(X-u/2)\right]^{n}$.
%%%%%%%%%%%%%%%%%%%%%%%%%%%%%%%%%%%%%%%%%%%%%%%%%%%%%%%%%%%%%%%%%
\hypertarget{S3B}{}\subsection{B. Connection to Hudson's theorem}\label{sec:S3B}
To further understand why it is crucial to properly account for the cavity response to generate Wigner negativity in the unresolved-sideband regime, let us consider the case of an initial pure state $\rhoin=\ket{\Psi}\bra{\Psi}$ with position wavefunction $\Psi_{i}(X)$. If $n$ photons are detected after the nonlinear cavity optomechanical interaction $U=\rme^{\rmi\varphi(X)\nl}$, then the final mechanical wavefunction is $\Psi_{n}(X)=\rme^{\rmi\varphi(X)n}\Psi_{i}(X)$. Whereas, if the full cavity response is not taken into account, $U=\rme^{\rmi\nl(\mu X+2\Db)}$ and the final mechanical wavefunction is $\Psi_{n}'(X)=\rme^{\rmi\mu n X}\Psi_{i}(X)$.

Hudson's theorem~\cite{hudson1974wigner} states that if a Wigner function is positive [$W(X,P)\geq0$ $\forall X,P$] its wavefunction must be of the form $\Psi(X)=\rme^{-aX^2+bX+c}$ for $a,b,c\in\mathbb{C}$ with $\mathrm{Re}(a)>0$. Importantly, the state $\Psi_{n}'(X)$ respects Hudson's theorem and so its Wigner function is positive. However, as $\varphi(X)$ cannot be written in the form $aX^2-bX-c$, $\Psi_{n}(X)$ possesses Wigner negativity.

%

%%%%%%%%%%%%%%%%%%%%%%%%%%%%%%%%%%%%%%%%%%%%%%%%%%%%%%%%%%%%%%%%%
\hypertarget{S3C}{}\subsection{C. Single-photon detection}\label{sec:S3C}

%%%%%%%%%%%%%%%%%%%%%%%%%%%%%%%%%%%%%%%%%%%%%%%%%%%%%%%%%%%%%%%%%
%%%%%%%%%%%%%%%%%% Appendix Figure Single Photon %%%%%%%%%%%%%%%%
%%%%%%%%%%%%%%%%%%%%%%%%%%%%%%%%%%%%%%%%%%%%%%%%%%%%%%%%%%%%%%%%%
\begin{figure*}
    \centering
    \includegraphics[width=\textwidth]{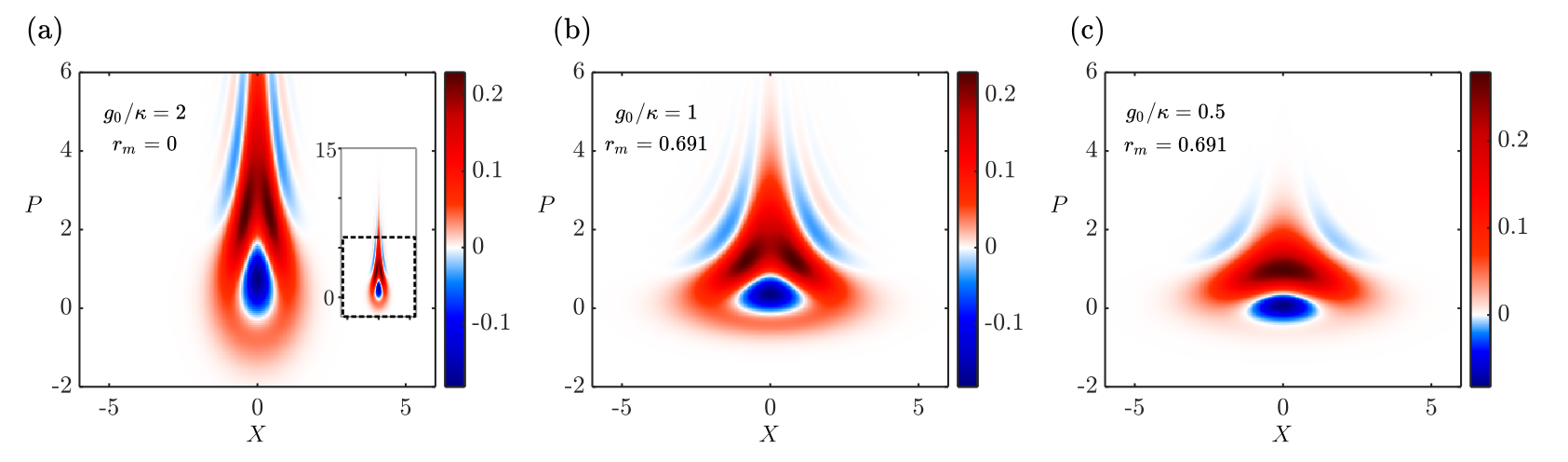}
    \caption{Enhancement of mechanical Wigner negativity via single-photon detection. (a) The Wigner function of a mechanical state generated from an initial mechanical ground state via an optomechanical interaction with a pulse of light. Here, $n=1$ photons are detected after the interaction and $g_{0}/\kappa=2$. The mechanical state does not depend on the state of the optical pulse and compared to the deterministic case [cf. Fig.~\ref{fig_1_main}(a)] $\delta$ increases to 0.39. (b) By squeezing the initial mechanical ground state (here, with a squeezing parameter of $r_{m}=0.691$) the same value of $\delta=0.39$ may be generated with a reduced coupling strength of $g_{0}/\kappa=1$. (c) Reducing the value of the coupling strength further to $g_{0}/\kappa=0.5$ decreases $\delta$ to 0.12. Compared to Fig.~\ref{fig_1_main}(c), optical squeezing does not provide a route to increase Wigner negativity as the final mechanical state is now insensitive to the quantum state of the optical pulse.}
    \label{appendix_fig_single_photon}
\end{figure*}
%%%%%%%%%%%%%%%%%%%%%%%%%%%%%%%%%%%%%%%%%%%%%%%%%%%%%%%%%%%%%%%%%

Here, we consider the particular case where one photon is detected following the nonlinear cavity optomechanical interaction ($n=1$). The input optical state may be any state that has $\Ph_{1}\neq0$, such as a coherent state $\ket{\alpha}$. In the case of $\ket{\psi}=\ket{\alpha}$,  the maximum value for the heralding probability is found at $|\alpha|=1$ and is given by $\mathrm{max}(\Ph_{1})=1/\rme$. Regardless of the form of $\ket{\psi}$, the Wigner function after one photon is detected is 
\begin{eqnarray}
W_{1}(X,P)&=&\frac{1}{2\pi}\int_{-\infty}^{+\infty}~\rme^{\rmi Pu}f^{*}(X+u/2)f(X-u/2)\bra{X-u/2}\rhoin\ket{X+u/2}~\rmd{u}.
\end{eqnarray}
For simplicity, we consider an initial mechanical state centred at the origin of phase space $X_{G}=P_{G}=0$. We may then write %
\begin{eqnarray}
\bra{X-u/2}\rhoin\ket{X+u/2}&=&W_{i}(X,P)\sqrt{\frac{d}{2\pi V_{X}}}g^{*}(u)
\end{eqnarray}
with $g(u)=\rme^{-\mathcal{A}(u+\rmi \mathcal{B})^2}$, $\mathcal{A}=d/2V_{X}$, and $\mathcal{B}=\left(V_{X}P-V_{XP}X\right)/d$. The Wigner function is then
\begin{eqnarray}
W_{1}(X,P)&=&W_{i}(X,P)\left[1+\sqrt{\frac{d}{2\pi V_{X}}}\int~h(u)g^{*}(u)~\rmd{u}\right]
\end{eqnarray}
with $h(u)=f^{*}(X+u/2)f(X-u/2)-1$, or, more explicitly,
\begin{equation}
    h(u)=\dfrac{-4\rmi\mu u}{\left[\frac{\mu}{2}\left(X-\frac{u}{2}\right)+\Db+\rmi\right]\left[\frac{\mu}{2}\left(X+\frac{u}{2}\right)+\Db-\rmi\right]}.
\end{equation}
Then noting the Fourier transform of $g(u)$ is $\tilde{g}(v)=\rme^{-v^2/2\mathcal{A}+\mathcal{B}v}/\sqrt{2\mathcal{A}}$ is a real function $\tilde{g}(v)=\tilde{g}^{*}(v)$, we may then use the Plancherel theorem to write the Wigner function as 
\begin{eqnarray}
W_{1}(X,P)&=&W_{i}(X,P)\left[1+\sqrt{\frac{d}{2\pi V_{X}}}\int~\tilde{h}(v)\tilde{g}(v)~\rmd{v}\right]\nonumber\\
&=&W_{i}(X,P)\left\{1-\dfrac{16\pi}{\mu(2\Db+\mu X)}\sqrt{\frac{d}{2\pi V_{X}}}\mathrm{Re}\left[y\rme^{z^2}\mathrm{erfc}(z)\right]\right\},
\end{eqnarray}
where $y=2\rmi+2\Db+\mu X$ and $z=\sqrt{\mathcal{A}}\left(4-4\rmi\Db-\mu \mathcal{B}-2\rmi \mu X\right)/\mu$.

%%%%%%%%%%%%%%%%%%%%%%%%%%%%%%%%%%%%%%%%%%%%%%%%%%%%%%%%%%%%%%%%%
\hypertarget{S4}{}\section{IV. Environmental effects}\label{sec:S4}

%%%%%%%%%%%%%%%%%%%%%%%%%%%%%%%%%%%%%%%%%%%%%%%%%%%%%%%%%%%%%%%%%
\hypertarget{S4A}{}\subsection{A. Optical loss}\label{sec:S4A}
%%%%%%%%%%%%%%%%%%%%%%%%%%%%%%%%%%%%%%%%%%%%%%%%%%%%%%%%%%%%%%%%%
\subsubsection{Deterministic case}
The mechanical state generated deterministically is unaffected by optical loss. To show this, consider a fictitious beamsplitter between the output optical mode and the optical environmental mode, which is assumed to be in the vacuum state $\ket{0}$. The beamsplitter is modelled by the unitary $B$ with transmission coefficient $\eta$. Hence, the state of the mechanical mode after the optomechanical interaction $U$ and optical loss is given by tracing over the optical mode of interest ($l$) and the optical environmental mode ($E$): 
\begin{eqnarray}
\rhof=\tr_{l,E}\left(B U \ket{\psi}\bra{\psi}\otimes\ket{0}\bra{0}\otimes\rhoin U^{\dagger} B^{\dagger}\right).
\end{eqnarray}
As the unitary $B$ acts only in the subspace of the two optical modes, we may use the cyclic property of the trace to rewrite $\rhof$ as
\begin{eqnarray}
\rhof&=&\tr_{l,E}\left(B^{\dagger}B U \ket{\psi}\bra{\psi}\otimes\ket{0}_{E}\bra{0}\otimes\rhoin U^{\dagger} \right)\nonumber\\
&=&\tr_{l,E}\left(U \ket{\psi}\bra{\psi}\otimes\ket{0}\bra{0}\otimes\rhoin U^{\dagger} \right)\nonumber\\
&=&\tr_{l}\left(U \ket{\psi}\bra{\psi}\otimes\rhoin U^{\dagger} \right),
\end{eqnarray}
which is the same as the mechanical state generated in the absence of optical loss.

%%%%%%%%%%%%%%%%%%%%%%%%%%%%%%%%%%%%%%%%%%%%%%%%%%%%%%%%%%%%%%%%%
\subsubsection{Probabilistic case}
However, if $n$-photons are detected after the optomechanical interaction, the mechanical state generated is affected by optical loss. This can be seen as follows. First, consider the optomechanical interaction $U$ between the initial pure optical state $\ket{\psi}$ and the initial mechanical state $\rhoin$. After the interaction, let the output optical mode interact via the fictitious beamsplitter $B$ with an environmental vacuum mode $\ket{0}$. If $n$ photons are detected by the photon-counting measurement, the final mechanical state is given by
\begin{eqnarray}
\rho_{n}=\dfrac{\sum_{m}\Upsilon_{n,m}\rhoin\Upsilon_{n,m}^{\dagger}}{\Ph_{n}},
\end{eqnarray}
where $\Ph_{n}=\tr\left(\Upsilon_{n,m}^{\dagger}\Upsilon_{n,m}\rhoin\right)$. Here, $m$ labels the number of photons detected by the optical environment, which are traced over, and the measurement operator $\Upsilon_{n,m}$ is given by
\begin{eqnarray}
\Upsilon_{n,m}&=&\bra{n,m}BU\ket{\psi}\otimes\ket{0}\nonumber\\
&=&\sum_{l=0}^{\infty}c_{l}\rme^{\rmi\varphi(X)l}\bra{n,m}B\ket{l,0}\nonumber\\
&=&\sum_{l=0}^{\infty}c_{l}\rme^{\rmi\varphi(X)l}\sum_{k=0}^{l}\sqrt{\begin{pmatrix}l\\k\end{pmatrix}}\left(\sqrt{\eta}\right)^{k}\left(\sqrt{1-\eta}\right)^{l-k}\bra{n,m}\ket{k,l-k}\nonumber\\
&=&\sum_{l=0}^{\infty}\sum_{k=0}^{l}c_{l}\delta_{n,k}\delta_{m,l-k}\rme^{\rmi\varphi(X)l}\sqrt{\begin{pmatrix}l\\k\end{pmatrix}}\left(\sqrt{\eta}\right)^{k}\left(\sqrt{1-\eta}\right)^{l-k}\nonumber\\
&=&\sum_{l=0}^{\infty}c_{l}\delta_{l,n+m}\rme^{\rmi\varphi(X)l}\sqrt{\begin{pmatrix}l\\n\end{pmatrix}}\left(\sqrt{\eta}\right)^{n}\left(\sqrt{1-\eta}\right)^{l-n}\nonumber\\
&=&c_{n+m}\rme^{\rmi\varphi(X)(n+m)}\sqrt{\begin{pmatrix}n+m\\n\end{pmatrix}}\left(\sqrt{\eta}\right)^{n}\left(\sqrt{1-\eta}\right)^{m}.
\end{eqnarray}
Hence, for an $n$-photon detection event the heralding probability is 
\begin{eqnarray}
\Ph_{n}&=&\sum_{m}|c_{n+m}|^2\begin{pmatrix}n+m\\n\end{pmatrix}\eta^{n}\left(1-\eta\right)^{m}\label{eq:heralding_prob_photon_det_loss},
\end{eqnarray} 
the mechanical density operator is
\begin{eqnarray}
    \rho_{n}=\dfrac{\sum_{m}|c_{n+m}|^2\begin{pmatrix}n+m\\n\end{pmatrix}\eta^{n}\left(1-\eta\right)^{m}\rme^{\rmi\varphi(X)(n+m)}\rhoin\rme^{-\rmi\varphi(X)(n+m)}}{\sum_{m}|c_{n+m}|^2\begin{pmatrix}n+m\\n\end{pmatrix}\eta^{n}\left(1-\eta\right)^{m}},
\end{eqnarray}
and the mechanical Wigner function is
\begin{eqnarray}
     W_{n}(X,P)&=&\frac{1}{2\pi}\int_{-\infty}^{+\infty}~\rme^{\rmi Pu}\mathcal{K}(X,u)~\bra{X-u/2}\rhoin\ket{X+u/2}~\rmd{u},\\
     \mathcal{K}(X,u)&=&\dfrac{\left[\eta f^{*}(X+u/2)f(X-u/2)\right]^{n}}{\Ph_{n}}\sum_{m}|c_{n+m}|^2\begin{pmatrix}n+m\\n\end{pmatrix}\left[(1-\eta) f^{*}(X+u/2)f(X-u/2)\right]^{m}.
\end{eqnarray}

For example, when $\ket{\psi}=\ket{\alpha}$, we have
\begin{eqnarray}
    \Ph_{n}&=&\rme^{-\eta|\alpha|^2}\dfrac{\left(\eta|\alpha|^2\right)^{n}}{n!},\\
    \rho_{n}&=&\rme^{\rmi\varphi(X)n}\left(\rme^{-(1-\eta)|\alpha|^2}\sum_{m=0}\dfrac{\left[(1-\eta)|\alpha|^2\right]^m}{m!}\rme^{\rmi\varphi(X)m}\rhoin\rme^{-\rmi\varphi(X)m}\right)\rme^{-\rmi\varphi(X)n},\\
    W_{n}(X,P)&=&\frac{1}{2\pi}\int_{-\infty}^{+\infty}~\rme^{\rmi Pu}\mathcal{K}(X,u)~\bra{X-u/2}\rhoin\ket{X+u/2}~\rmd{u},\\
    \mathcal{K}(X,u)&=&\left[f^{*}(X+u/2)f(X-u/2)\right]^{n}\exp\left\{-(1-\eta)|\alpha|^2\dfrac{\rmi\mu u}{\left[\frac{\mu}{2}(X+\frac{u}{2})+\Db-\rmi\right]\left[\frac{\mu}{2}(X-\frac{u}{2})+\Db+\rmi\right]}\right\}.
\end{eqnarray}
When there is no optical loss, $\eta=1$, we recover the Eq.~\eqref{eqsupp:W_n}, and when $\eta=0$, which corresponds to a trace operation, the only value of $n$ that occurs with a non-zero heralding probability is $n=0$ and hence we recover the deterministic result of Eq.~\eqref{eqsupp:Wf_alpha}.

%%%%%%%%%%%%%%%%%%%%%%%%%%%%%%%%%%%%%%%%%%%%%%%%%%%%%%%%%%%%%%%%%
\hypertarget{S4B}{}\subsection{B. Mechanical thermal effects \& nonclassical depth}\label{sec:S4B}

%%%%%%%%%%%%%%%%%%%%%%%%%%%%%%%%%%%%%%%%%%%%%%%%%%%%%%%%%%%%%%%%%
%%%%%%%%%%%%%%%%%%%%%%%%%%%% Figure Tau %%%%%%%%%%%%%%%%%%%%%%%%%
%%%%%%%%%%%%%%%%%%%%%%%%%%%%%%%%%%%%%%%%%%%%%%%%%%%%%%%%%%%%%%%%%
\begin{figure*}
    \centering
    \includegraphics[width=0.5\textwidth]{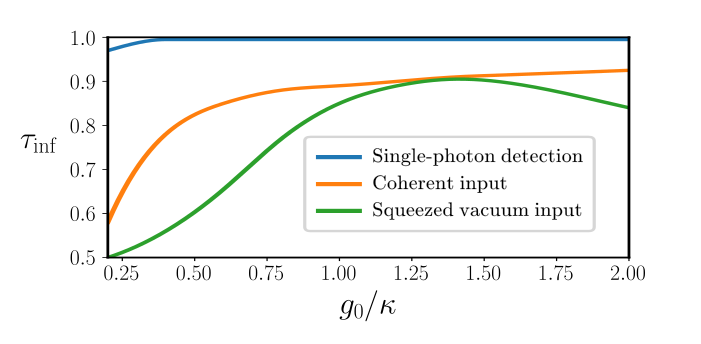}
    \caption{ The nonclassical depth $\tau_{\mathrm{inf}}$ for a mechanical state created with a coherent input ($\alpha=2$), a squeezed vacuum input ($r_{l}=0.691$), and via single-photon detection. Here, $\Db=0$, $r_{m}=0.691$, and $\bar{N}=0$.}
    \label{appendix_fig_tau_inf}
\end{figure*}

Between state generation and verification, the addition of $\tauth$ thermal phonons from the mechanical environment can be modelled via the map
\begin{eqnarray}\label{eq:thermal_map_state}
    \rho_{\mathrm{th}}=\frac{1}{\pi\tauth}\int~\rme^{-|\beta'|^2/\tauth}D(\beta')\rho D^{\dagger}(\beta')~\rmd^2 \beta',
\end{eqnarray}
where $\beta=(X'+\rmi P')/\sqrt{2}$~[\href{https://doi.org/10.1103/PhysRevA.51.4967}{Z. H. Musslimani, S. L. Braunstein, A. Mann, and M. Revzen, Phys. Rev. A \textbf{51}, 4967 (1995)}]. To derive how Eq.~\eqref{eq:thermal_map_state} changes the Wigner function of the mechanical state, we use the definition
\begin{eqnarray}
    W(\beta)=\frac{1}{\pi^2}\int~\rme^{\beta\lambda^*-\beta^*\lambda}\tr\left[\rho D(\lambda)\right]~\rmd^2 \lambda,
\end{eqnarray}
and the braiding relation $D(\lambda)D(\beta)=D(\beta)D(\lambda)\rme^{\lambda\beta^*-\lambda^*\beta}$. Thus, the Wigner-function transformation which corresponds to Eq.~\eqref{eq:thermal_map_state} is
\begin{eqnarray}\label{eq:thermal_map_Wig}
    W_{\mathrm{th}}(\beta)=\frac{1}{\pi\tauth}\int~\rme^{-|\beta'|^2/\tauth}W(\beta-\beta')~\rmd^2 \beta'.
\end{eqnarray}
Notably, an $s$-parametrized Wigner function, with $s<0$, can be defined as
\begin{eqnarray}\label{eq:sparam_Wig_1}
    W_{s}(\beta)=\frac{2}{\pi|s|}\int~\rme^{-2|\beta'|^2/|s|}W(\beta-\beta')~\rmd^2 \beta',
\end{eqnarray}
which is identical to Eq.~\eqref{eq:thermal_map_Wig} with $\tauth=|s|/2$. 

Likewise, an $s$-parametrized Wigner function, with $s<1$, can be defined in terms of the Glauber-Sudarshan $P$ function via
\begin{eqnarray}\label{eq:sparam_Wig_2}
    W_{s}(\beta)=\frac{2}{\pi(1-s)}\int~\rme^{-2|\beta-\beta'|^2/(1-s)}P(\beta')~\rmd^2 \beta',
\end{eqnarray}
which is equivalent to the $R$ function
\begin{eqnarray}\label{eq:Rfunc}
    R_{\tau}=\frac{1}{\pi \tau} \int~\rme^{-|\beta-\beta'|^2/\tau}P(\beta')~\rmd^2 \beta'
\end{eqnarray}
when $\tau=(1-s)/2$. The nonclassical depth $\tau_{\mathrm{inf}}$ of a state is defined as the  infimum over all values of $\tau$ for which the $R$ function is a valid classical probability distribution~\href{https://doi.org/10.1103/PhysRevA.44.R2775}{[C. T. Lee, Phys. Rev. A \textbf{44}, R2775(R) (1991)]}. To be more precise, we follow Ref.~\href{https://doi.org/10.1103/PhysRevA.93.053818}{[T. J. Milburn, M. S. Kim, and M. R. Vanner, Phys. Rev. A \textbf{93}, 053818 (2016)]} and define an acceptable classical probability distribution as one which is both non-negative and integrable. Thus, the nonclassical depth of a state is defined from Eq.~\eqref{eq:Rfunc} via 
\begin{eqnarray}
    \tau_{\mathrm{inf}}=\mathrm{inf}\left\{\tau:~R_{\tau}\text{ is non-negative and integrable}\right\}.
\end{eqnarray}

Starting from an initially negative Wigner function, such that the thermalized state is represented by $s<0$, we may then combine the relations $\tau=1/2+|s|/2$ and $\tauth=|s|/2$ to $\tau=1/2+\tauth$ to calculate the nonclassical depth numerically.

Fig.~\ref{appendix_fig_tau_inf} plots the nonclassical depth of the mechanical states generated deterministically and via single-photon detection. Here, we confirm that photon counting provides a feasible route to both enhance and verify mechanical negativity probabilistically. For the mechanical states generated deterministically, Fig.~\ref{appendix_fig_tau_inf} shows that the nonclassical depth generated via the coherent input pulse is larger than that of the mechanical state generated via the squeezed vacuum state. 
As the squeezed vacuum state does not contain the odd-numbered photon Fock states, for a given value of $g_{0}/\kappa$ the population components of the mechanical states are separated by a larger distance in phase space and are thus more susceptible to decoherence, which implies a lower value of $\tau_{\mathrm{inf}}$---note the negative volume indicating that these states are more nonclassical. The decrease in $\tau_{\mathrm{inf}}$ beyond $g_{0}/\kappa\simeq1.4$ for the squeezed vacuum input occurs as the increased rate of decoherence outweighs any increase of mechanical negativity with $g_{0}/\kappa$.

%%%%%%%%%%%%%%%%%%%%%%%%%%%%%%%%%%%%%%%%%%%%%%%%%%%%%%%%%%%%%%%%%
\hypertarget{S5}{}\section{V. Continuously-driven optomechanics}\label{sec:S5}

%%%%%%%%%%%%%%%%%%%%%%%%%%%%%%%%%%%%%%%%%%%%%%%%%%%%%%%%%%%%%%%%%
\hypertarget{S5A}{}\subsection{A. General Gaussian input optical state}\label{sec:S5A}
For the input optical mode, we consider an arbitrary pure single-mode Gaussian state $\ket{\psi_{G}}$, which may be generated by applying the rotation operator $R(\phi)=\rme^{\rmi\phi \nl}$, the displacement operator $D(\alpha)=\rme^{\alpha a^{\dagger}-\alpha^{*}a}$, and the squeezing operator $S(\zeta)=\rme^{\frac{1}{2}(\zeta^{*} a^2-\zeta a^{\dagger 2})}$ to the vacuum state $\ket{0}$, i.e. $\ket{\psi_{G}}=R(\phi)D(\alpha)S(\zeta)\ket{0}$.
Here, $0\leq\phi<2\pi$, $\alpha\in\mathbb{C}$, and $\zeta=r\rme^{\rmi\theta}$, where $r\geq0$ and $0\leq\theta<2\pi$. In the following, it will be useful to write the general Gaussian state as a rotation operation acting on the squeezed coherent state $\ket{\alpha,\zeta}$, i.e. $\ket{\psi_{G}}=R(\phi)\ket{\alpha,\zeta}$, where
\begin{equation}
    \ket{\alpha,\zeta}=\sum_{n}^{\infty}c_{n}\ket{n},
\end{equation}
\begin{equation}
      c_{n}=\dfrac{1}{\sqrt{\cosh r}}\exp\left[-\frac{1}{2}|\alpha|^2-\frac{1}{2}\alpha^{* 2}\rme^{\rmi \theta}\tanh r\right]\dfrac{\left(\frac{1}{2}\rme^{\rmi\theta}\tanh r\right)^{n/2}}{\sqrt{n!}}H_{n}\left[\dfrac{\alpha\cosh r+\alpha^{*}\rme^{\rmi\theta}\sinh r}{\sqrt{\rme^{\rmi\theta}\sinh 2r}}\right],
\end{equation}
and $H_{n}$ is the $n^{\mathrm{th}}$ Hermite polynomial~[{\href{https://www.google.co.uk/books/edition/Introductory_Quantum_Optics/CgByyoBJJwgC?hl=en&gbpv=0}{C. C Gerry and P. L. Knight, \emph{Introductory quantum optics} (Cambridge University Press, 2005)}}]. The total photon number in the input optical state is given by
\begin{eqnarray}
    \expval{\nl}&=&\bra{\psi_{G}}a^{\dagger}a\ket{\psi_{G}}\\
    &=&|\alpha|^2+\sinh^{2}r.
\end{eqnarray}
 A constant input photon flux $\expval{\nl}/\Delta{t}$ can also be defined over a small time increment $\Delta{t}$. To do this, we write 
\begin{equation}
  \alpha=\alphain\sqrt{\Delta{t}}\label{eq:alpha_flux}   
\end{equation}
and 
\begin{equation}
 r=\rin\sqrt{\Delta{t}},\label{eq:r_flux}
\end{equation}
 which for small $\Delta{t}$ gives 
\begin{equation}
    \expval{\nl}=\left(|\alphain|^2+\rin^2\right)\Delta{t}.
\end{equation}
Hence, the constant input photon flux is given by $\expval{\nl}/\Delta{t}=\left(|\alphain|^2+\rin^2\right)$. 

%%%%%%%%%%%%%%%%%%%%%%%%%%%%%%%%%%%%%%%%%%%%%%%%%%%%%%%%%%%%%%%%%
\hypertarget{S5B}{}\subsection{B. The master equation}\label{sec:S5B}

If the optical mode is not measured after the optomechanical interaction, the evolution of the mechanical mode is calculated by tracing over all possible measurement outcomes. Performing this trace operation in the number basis gives
\begin{equation}
    \rhof=\sum_{n}^{\infty}\Upsilon_{n}\rhoin\Upsilon^{\dagger}_{n},
\end{equation}
where the measurement operator for an $n$-photon detection is
\begin{eqnarray}
    \Upsilon_{n}&=&\bra{n}U\ket{\psi_{G}}\\
    &=&\bra{n}UR\ket{\alpha,\zeta}\\
    &=&c_{n}\rme^{\rmi\theta n}\rme^{\rmi\varphi(X)n}.
\end{eqnarray}
Hence, $\rhof$ may be written as 
\begin{equation}
    \rhof=\sum_{n}^{\infty}|c_{n}|^2\rme^{\rmi\varphi(X) n}\rhoin \rme^{-\rmi\varphi(X) n},\label{eq:measurement_map_continuous_1}
\end{equation}
which does not depend on the rotation angle $\phi$ of the general Gaussian state $\ket{\psi_{G}}$.

To derive the master equation that describes the continuous evolution of the mechanical mode, we consider the measurement map of Eq.~\eqref{eq:measurement_map_continuous_1} over a time interval $t\rightarrow t+\rmd{t}$. In other words, in Eq.~\eqref{eq:measurement_map_continuous_1} we let $\rhof=\rho(t+\rmd{t})$, $\rhoin=\rho(\rmd{t})$ and expand each of the coefficients $|c_{n}|^2$ to first order in $\rmd{t}$. The coefficients $|c_{n}|^2$ may be expanded to first order in $\rmd{t}$ by using Eqs.~\eqref{eq:alpha_flux} and \eqref{eq:r_flux}, taking the limit $\Delta{t}\rightarrow\rmd{t}$, and neglecting terms of order $\rmd{t}^2$ or higher, which gives
\begin{eqnarray}
    |c_{0}|^2&=&1-\left(|\alphain|^2+\frac{1}{2}\rin^2\right)\rmd{t},\\
    |c_{1}|^2&=&|\alphain|^2\rmd{t},\\
    |c_{2}|^2&=&\frac{1}{2}\rin^2\rmd{t},\\
    |c_{n}|^2&=&0\quad\text{for }n>2.
\end{eqnarray}
Inserting the expanded coefficients $|c_{n}|^2$ into Eq.~\eqref{eq:measurement_map_continuous_1} then gives
\begin{eqnarray}
    \rho(t+\rmd{t})=\rho(t)\left[1-\left(|\alphain|^2+\frac{1}{2}\rin^2\right)\rmd{t}\right]+|\alphain|^2\rmd{t}\rme^{\rmi\varphi(X)}\rho(t)\rme^{-\rmi\varphi(X)}+\frac{1}{2}|\rin|^2\rmd{t}\rme^{2\rmi\varphi(X)}\rho(t)\rme^{-2\rmi\varphi(X)}
\end{eqnarray}
Identifying $\rmd{\rho}$ as $\rho(t+\rmd{t})-\rho(t)$, and including the free evolution of the mechanical mode $H_{0}=\hbar\wm b^{\dagger}b$ and mechanical interactions with the thermal bath, the master equation for the mechanical mode interacting with a general Gaussian optical input state $\ket{\psi_{G}}$ is
\begin{equation}
    \dot{\rho}=-\frac{\rmi}{\hbar}[H_{0},\rho]+2\gamma\left(\bar{N}+1\right)\mathcal{D}\left[b\right]\rho+2\gamma\bar{N}\mathcal{D}\left[b^{\dagger}\right]\rho+\mathcal{D}\left[L_{1}\right]\rho+\mathcal{D}\left[L_{2}\right]\rho,\label{eq:ME}
\end{equation}
where $\gamma$ is the mechanical decay rate, $\bar{N}$ is the occupation of the thermal environment, $L_{n}=\sqrt{2k_{n}}\rme^{n\rmi\varphi(X)}=\sqrt{2k_{n}}f^n(X)$ for $n=1, 2$, $k_{1}=\frac{1}{2}|\alphain|^2$, and $k_{2}=\frac{1}{4}|\rin|^2$. The Lindblad superoperator is defined by $\mathcal{D}\left[O\right]:=O\rho O^{\dagger}-\frac{1}{2}\left\{O^{\dagger}O,\rho\right\}$.

%%%%%%%%%%%%%%%%%%%%%%%%%%%%%%%%%%%%%%%%%%%%%%%%%%%%%%%%%%%%%%%%%
\hypertarget{S5C}{}\subsection{C. Steady state \& the rotating wave approximation}\label{sec:S5C}
To find a steady-state solution for a continuously driven optomechanical system, we consider the case where $\Db=k_{2}=0$ and $k_{1}=k$. In this case, in a frame rotating at the mechanical frequency $\wm$, Eq.~\eqref{eq:ME} becomes 
\begin{equation}
    \dot{\rho}=2\gamma\left(\bar{N}+1\right)\mathcal{D}\left[b\right]\rho+2\gamma\bar{N}\mathcal{D}\left[b^{\dagger}\right]\rho+2k\mathcal{D}\left[\tilde{f}\right]\rho,\label{eq:ME_RWA}
\end{equation}
where $\tilde{f}=f(\tilde{X})=\left(1+\rmi\frac{\mu}{2}{\tilde{X}}\right)/\left(1-\rmi\frac{\mu}{2}{\tilde{X}}\right)$ and $\tilde{X}=(b\rme^{-\rmi\wm t}+b^{\dagger}\rme^{\rmi\wm t})/\sqrt{2}$. For convenience, we define $T=1-\rmi\frac{\mu}{2}{\tilde{X}}$ and so $\tilde{f}=T^{\dagger}/T=T^{\dagger}T^{-1}=T^{-1}T^{\dagger}$.

The free mechanical evolution and open-system dynamics are phase-symmetric operations, i.e. $[H_{0},\bd b]=0$, $\mathcal{D}\left[b\rme^{\rmi\theta}\right]\rho=\mathcal{D}\left[b\right]\rho$, and $\mathcal{D}\left[\bd\rme^{\rmi\theta}\right]\rho=\mathcal{D}\left[\bd\right]\rho$. However, the term describing the continuous optomechanical interaction is not phase symmetric. Thus, to find an approximate steady-state solution to Eq.~\eqref{eq:ME_RWA} we make a rotating wave approximation (RWA) on the term $\mathcal{D}[\tilde{f}]\rho=\tilde{f}\rho\tilde{f}^{\dagger}-\rho$. When the RWA is valid, steady-state solutions to Eq.~\eqref{eq:ME_RWA}  will therefore be a general phase-symmetric state of the form $\rho=\sum_{n=0}^{\infty}~P_{n}\ket{n}\bra{n}$. To proceed, we consider the action of $\tilde{f}=T^{-1}T^{\dagger}$ on an arbitrary Fock state $\ket{n}$.

%%%%%%%%%%%%%%%%%%%%%%%%%%%%%%%%%%%%%%%%%%%%%%%%%%%%%%%%%%%%%%%%%
\subsubsection{Action of $\tilde{f}$ on a Fock state}

Firstly, we note that $T^{\dagger}\ket{n}=\ket{n}+\rmi\frac{\mu}{2\sqrt{2}}\sqrt{n}\rme^{-\rmi\wm t}\ket{n-1}+\rmi\frac{\mu}{2\sqrt{2}}\sqrt{n+1}\rme^{\rmi\wm t}\ket{n+1}$ for $n>0$ and $T^{\dagger}\ket{n}=\ket{0}+\rmi\frac{\mu}{2\sqrt{2}}\rme^{\rmi\wm t}\ket{1}$ for $n=0$. We write these expression as
\begin{equation}
    T^{\dagger}\ket{n}=\sum_{k=n-1}^{n+1}~d_{k+1,n+1}\rme^{\rmi(k-n)\wm t}\ket{k},\label{eq:Tdag_on_n}
\end{equation}
where the coefficients $d_{k+1,n+1}$ are defined via
\begin{eqnarray}
    d_{n,n+1}&=&\rmi\frac{\mu}{2\sqrt{2}}\sqrt{n},\\
    d_{n+1,n+1}&=&1,\\
    d_{n+2,n+1}&=&\rmi\frac{\mu}{2\sqrt{2}}\sqrt{n+1},\\
    d_{k+1,n+1}&=&0\quad\text{for }k\neq n,n+1,n+2,
\end{eqnarray}
and also $d_{0,1}=0$. Secondly, we turn to the inverse of the operator $T=1-\rmi\frac{\mu}{2}{\tilde{X}}$. To this end, we represent $T$ as an $N\times N$ matrix in the Fock basis as
\begin{equation}
    T=\begin{pmatrix}
        a_{1}  & b_{1}  & 0      & 0      & \dots   & 0       \\
        c_{1}  & a_{2}  & b_{2}  & 0      & \dots   & 0       \\
        0      & c_{2}  & a_{3}  & b_{3}  & \dots   & 0       \\
        0      & 0      & c_{3}  & \ddots & \ddots  & 0       \\
        \vdots & \vdots & \vdots & \ddots & \ddots  & b_{N-1} \\
        0      & 0      & 0      & 0      & c_{N-1} & a_{N}
    \end{pmatrix}
\end{equation}
with
\begin{eqnarray}
    a_{j}&=&1,\\
    b_{j}&=&-\rmi\frac{\mu}{2\sqrt{2}}\sqrt{j}\rme^{-\rmi\wm t},\\
    c_{j}&=&-\rmi\frac{\mu}{2\sqrt{2}}\sqrt{j}\rme^{\rmi\wm t},
\end{eqnarray}
and all other elements are equal to zero. Note, that we truncate the Fock space so that $T$ is an $N\times N$ matrix and $N$ is chosen to be sufficiently high such that further increasing $N$ has negligible effect on the final mechanical state. Thus, in the Fock basis $T$ takes the form of a $N \times N$ tridiagonal matrix. Its inverse may be calculated following Ref.~[\href{https://doi.org/10.1016/0898-1221(94)90066-3}{R.~A. Usmani, Comp. Math. Applic. \textbf{27}, 59 (1994)}] by first considering the following recurrence relations:
\begin{eqnarray}
    \theta_{j}&=&a_{j}\theta_{j-1}-b_{j-1}c_{j-1}\theta_{j-2}\nonumber\\
    &=&\theta_{j-1}+(j-1)\frac{\mu^2}{8}\theta_{j-2},\label{eq:theta_rec}
\end{eqnarray}
for $j=2,3,\ldots,N$ with initial conditions $\theta_{0}=\theta_{1}=1$ and
\begin{eqnarray}
    \phi_{j}&=&a_{j}\phi_{j+1}-b_{j}c_{j}\phi_{j+2}\nonumber\\
    &=&\phi_{j+1}+j\frac{\mu^2}{8}\phi_{j+2},\label{eq:phi_rec}
\end{eqnarray}
for $j=N-1,N-2,\ldots,1$ with initial conditions $\phi_{N+1}=\phi_{N}=1$. Note that Eqs~\eqref{eq:theta_rec} and \eqref{eq:phi_rec} are independent of the mechanical frequency and $\theta_{N}=\mathrm{det}(T)$. With these recurrence relations, the elements of $T^{-1}$ can then be calculated:
\begin{eqnarray}
    T^{-1}_{ij}&=&\begin{cases}
                    & (-1)^{i+j}b_{i}b_{i+1}\ldots b_{j-1}\dfrac{\theta_{i-1}\phi_{j+1}}{\theta_{N}} ~~~\text{ for } i<j,\\
                    & \dfrac{\theta_{i-1}\phi_{j+1}}{\theta_{N}} \qquad\text{ for } i=j,\\
                    & (-1)^{i+j}c_{j}c_{j+1}\ldots c_{i-1}\dfrac{\theta_{j-1}\phi_{i+1}}{\theta_{N}} ~~~\text{ for } i>j.
                 \end{cases}\nonumber\\
\end{eqnarray}
Further, by using the expressions for $b_{j}$ and $c_{j}$, one finds
\begin{eqnarray}
    T^{-1}_{ij}&=&\begin{cases}
                    & (-1)^{i+j}\left(-\rmi\frac{\mu}{2\sqrt{2}}\rme^{-\rmi\wm t}\right)^{j-i}\sqrt{\dfrac{(j-1)!}{(i-1)!}}\dfrac{\theta_{i-1}\phi_{j+1}}{\theta_{N}} ~~~\text{ for } i<j,\\
                    & \dfrac{\theta_{i-1}\phi_{j+1}}{\theta_{N}} \qquad\text{ for } i=j,\\
                    & (-1)^{i+j}\left(-\rmi\frac{\mu}{2\sqrt{2}}\rme^{\rmi\wm t}\right)^{i-j}\sqrt{\dfrac{(i-1)!}{(j-1)!}}\dfrac{\theta_{j-1}\phi_{i+1}}{\theta_{N}} ~~~~~\text{ for } i>j.
                 \end{cases}   
\end{eqnarray}
We then write these elements as $T^{-1}_{ij}=C_{ij}\rme^{(i-j)\rmi\wm t}$, where 
\begin{eqnarray}
    C_{ij}&=&\begin{cases}
                    & (-1)^{i+j}\left(-\rmi\frac{\mu}{2\sqrt{2}}\right)^{j-i}\sqrt{\dfrac{(j-1)!}{(i-1)!}}\dfrac{\theta_{i-1}\phi_{j+1}}{\theta_{N}} ~~~\text{ for } i<j,\\
                    & \dfrac{\theta_{i-1}\phi_{j+1}}{\theta_{N}} \qquad\text{ for } i=j,\\
                    & (-1)^{i+j}\left(-\rmi\frac{\mu}{2\sqrt{2}}\right)^{i-j}\sqrt{\dfrac{(i-1)!}{(j-1)!}}\dfrac{\theta_{j-1}\phi_{i+1}}{\theta_{N}} ~~~\text{ for } i>j,
                 \end{cases}   
\end{eqnarray}
which allows for a similar expression to Eq.~\eqref{eq:Tdag_on_n} to be written:
\begin{equation}
    T^{-1}\ket{n}=\sum_{m=0}^{N-1}~C_{m+1,n+1}\rme^{\rmi(m-n)\wm t}\ket{m}.\label{eq:Tinv_on_n}
\end{equation}
Together, Eqs~\eqref{eq:Tdag_on_n} and \eqref{eq:Tinv_on_n} give the action on $\tilde{f}$ on a Fock state as
\begin{equation}
    \tilde{f}\ket{n}=\sum_{m=0}^{N-1}\sum_{k=m-1}^{m+1}~C_{m+1,n+1}d_{k+1,m+1}\rme^{\rmi(k-n)\wm t}\ket{k}.\label{eq:f_on_n}
\end{equation}
Note as the Fock space is truncated from $n=0$ to $n=N-1$, $\tilde{f}$ is also an $N\times N$ matrix.

\subsubsection{Rotating wave approximation}
To perform the RWA, consider the matrix element $\tilde{f}\ket{n}\bra{n}\tilde{f}^{\dagger}$. By using Eq.~\eqref{eq:f_on_n} this matrix element is
\begin{equation}
    \tilde{f}\ket{n}\bra{n}\tilde{f}^{\dagger}=\sum_{m=0}^{N-1}\sum_{k=m-1}^{m+1}\sum_{l=0}^{N-1}\sum_{p=l-1}^{l+1}~C_{m+1,n+1}d_{k+1,m+1}C^*_{l+1,n+1}d^*_{p+1,l+1}\rme^{\rmi(k-p)\wm t}\ket{k}\bra{p}\label{eq:matrix_element_nn}
\end{equation}
When the oscillating terms in Eq.~\eqref{eq:matrix_element_nn} are negligible, the matrix element is given by
\begin{eqnarray}
    \tilde{f}\ket{n}\bra{n}\tilde{f}^{\dagger}&=&\sum_{m=0}^{N-1}\sum_{q=-1}^{+1}\sum_{k=m-1}^{m+1}F_{n,m,q,k}\ket{m+q}\bra{m+q},\label{eq:RWA_element}\\
    F_{n,m,q,k}&:=&C_{m+1,n+1}d_{m+1+q,m+1}C^*_{k+1+q,n+1}d^*_{m+1+q,k+1+q}.
\end{eqnarray}
As the Fock space has been truncated, $C_{ij}$ and $d_{ij}$ are set to zero if $i,j\leq 0$ or if $i,j\geq N+1$.

\begin{figure*}
    \centering
    \includegraphics[width=\textwidth]{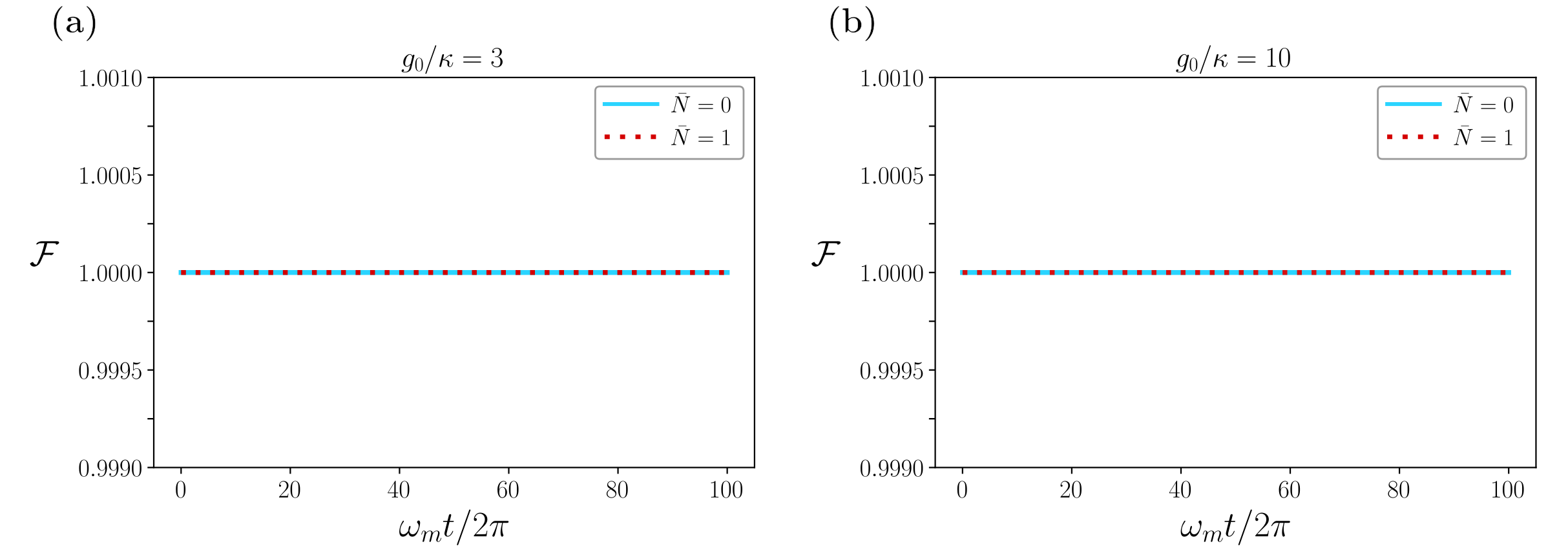}
    \caption{Testing the validity of the RWA numerically for (a) $g_{0}/\kappa$=3 and (b) $g_{0}/\kappa=10$. Here, we plot the quantum state fidelity $\mathcal{F}$ of the RWA solution with the numerical solution obtained using the \texttt{mesolve} function in QuTiP. The fidelity between the RWA and the numerical solution is equal to 1 over the 100 mechanical periods tested. Here, the parameters used are the same as those in Fig.~\ref{fig_5_main}, $k=0.1\,\mathrm{s}^{-1}$, and there are $100$ levels in the Hilbert space.   
    }
    \label{appendix_fig_fidelity}
\end{figure*}

Approximating Eq.~\eqref{eq:matrix_element_nn} by Eq.~\eqref{eq:RWA_element} is valid for the parameters explored in this work. To test the validity of the RWA, we use the RWA solution as an initial condition and numerically simulate the master equation (without making a RWA) for 100 mechanical periods using the QuTiP Python package and the inbuilt master equation solver: \texttt{mesolve}~[\href{https://doi.org/10.1016/j.cpc.2012.02.021}{J. R. Johansson, P. D. Nation, and F. Nori, Comput. Phys.
Commun. 183, 1760 (2012)}]. In Fig.~\ref{appendix_fig_fidelity}, we then plot the quantum state fidelity $\mathcal{F}$ between the RWA solution and the mechanical state acquired numerically using QuTiP. We see that $\mathcal{F}=1$ and the parameters used are the same as those in Fig.~\ref{fig_5_main} with $k=0.1\,\mathrm{s}^{-1}$. 

%%%%%%%%%%%%%%%%%%%%%%%%%%%%%%%%%%%%%%%%%%%%%%%%%%%%%%%%%%%%%%%%%
%%%%%%%%%%%%%%%%%%%%% Appendix Figure Fidelity %%%%%%%%%%%%%%%%%%
%%%%%%%%%%%%%%%%%%%%%%%%%%%%%%%%%%%%%%%%%%%%%%%%%%%%%%%%%%%%%%%%%
%%%%%%%%%%%%%%%%%%%%%%%%%%%%%%%%%%%%%%%%%%%%%%%%%%%%%%%%%%%%%%%%%
%%%%%%%%%%%%%%%%%%%%%%%%%%%%%%%%%%%%%%%%%%%%%%%%%%%%%%%%%%%%%%%%%

\subsubsection{The mechanical density matrix}

When the system operates in a parameter regime in which the RWA is valid, the steady-state mechanical density matrix will be of the form $\rho=\sum_{n=0}^{N-1}P_{n}\ket{n}\bra{n}$ as discussed above. To find the set of coefficients $\{P_{n}\}$, we calculate the matrix element $\bra{l}\dot{\rho}\ket{l}=\dot{\rho}_{ll}$ using Eq.~\eqref{eq:ME_RWA} and then set $\dot{\rho}_{ll}=0$. By denoting $\bra{l}\mathcal{D}\left[O\right]\rho\ket{l}$ as $\left\{\mathcal{D}\left[O\right]\rho\right\}_{ll}$, this procedure gives
\begin{eqnarray}
    \left\{\mathcal{D}\left[b\right]\rho\right\}_{ll}&=&\begin{cases}
                    & (l+1)P_{l+1}-lP_{l}  ~~~\text{ for } l>0,\\
                    & P_{1} ~~~\,\qquad\qquad\qquad\text{ for } l=0,
                 \end{cases}   \label{eq:Db_ll}
\end{eqnarray}
and similarly 
\begin{eqnarray}
    \left\{\mathcal{D}\left[b^{\dagger}\right]\rho\right\}_{ll}&=&\begin{cases}
                    & lP_{l-1}-(l+1)P_{l}  ~~~\text{ for } l>0,\\
                    & -P_{0} ~\,\qquad\qquad\qquad\text{ for } l=0.
                 \end{cases}  \label{eq:Dbdag_ll} 
\end{eqnarray}
In the RWA, the term corresponding to the continuous optomechanical interaction gives the relation
\begin{equation}
    \left\{\mathcal{D}\left[\tilde{f}\right]\rho\right\}_{ll}=\sum_{n=0}^{N-1}P_{n}\Phi_{n,l}-P_{l},\label{eq:omint_ll}
\end{equation}
where Eq.~\eqref{eq:RWA_element} was used and
\begin{equation}
    \Phi_{n,l}=\sum_{k=l}^{l+2}F_{n,l+1,-1,k}+\sum_{k=l-1}^{l+1}F_{n,l,0,k}+\sum_{k=l-2}^{l}F_{n,l-1,1,k},
\end{equation}
for $l>0$, and
\begin{equation}
    \Phi_{n,0}=\sum_{k=1}^{2}F_{n,1,-1,k}+\sum_{k=0}^{1}F_{n,0,0,k},
\end{equation}
for $l=0$. Inserting Eqs~\eqref{eq:Db_ll} to \eqref{eq:omint_ll} into Eq.~\eqref{eq:ME_RWA} and setting $\dot{\rho}_{ll}=0$ gives the linear system of coupled equations:
\begin{equation}
    0=2\gamma(\bar{N}+1)\left\{\mathcal{D}\left[b\right]\rho\right\}_{ll}+2\gamma\bar{N}\left\{\mathcal{D}\left[b^{\dagger}\right]\rho\right\}_{ll}+2k\left\{\mathcal{D}\left[\tilde{f}\right]\rho\right\}_{ll},\label{eq:system_of_eqs_1}
\end{equation}
where $l=0,1,\ldots,N-1$. To find the steady-state mechanical density matrix, the system of equations given by Eq.~\eqref{eq:system_of_eqs_1} must be solved subject to the condition that $\mathrm{tr}(\rho)=1$. When the Fock space is truncated at a sufficiently high value $N$, such that as $l$ approaches $N$ the probabilities $P_{l}$ are close to zero, the system of linear equations may be extended by including the normalization condition $\mathrm{tr}(\rho)=\sum_{n}^{N}P_{n}=1$. Thus, the set of coefficients $\{P_{n}\}$ is determined by solving the linear matrix equation $A\mathbf{P}=\mathrm{b}$. Here, $A$ is an $(N+1)\times(N+1)$ matrix, whose first $N$ rows correspond to the system of equations described by Eq.~\eqref{eq:system_of_eqs_1} and whose final row corresponds to the normalization condition. Note that to accommodate the normalization condition, $A_{i,N+1}=0$ for all $i\leq N$. Further, $\mathbf{P}=\left(P_{0},P_{1},\ldots,P_{N-1},P_{N}\right)^{\mathrm{T}}$ and $\mathbf{b}=\left(0,0,\ldots,0,1\right)^{\mathrm{T}}$.

%%%%%%%%%%%%%%%%%%%%%%%%%%%%%%%%%%%%%%%%%%%%%%%%%%%%%%%%%%%%%%%%%
%%%%%%%%%%%%%%%%%%%%% Appendix Figure delta %%%%%%%%%%%%%%%%%%%%%
%%%%%%%%%%%%%%%%%%%%%%%%%%%%%%%%%%%%%%%%%%%%%%%%%%%%%%%%%%%%%%%%%
\begin{figure*}
    \centering
    \includegraphics[width=\textwidth]{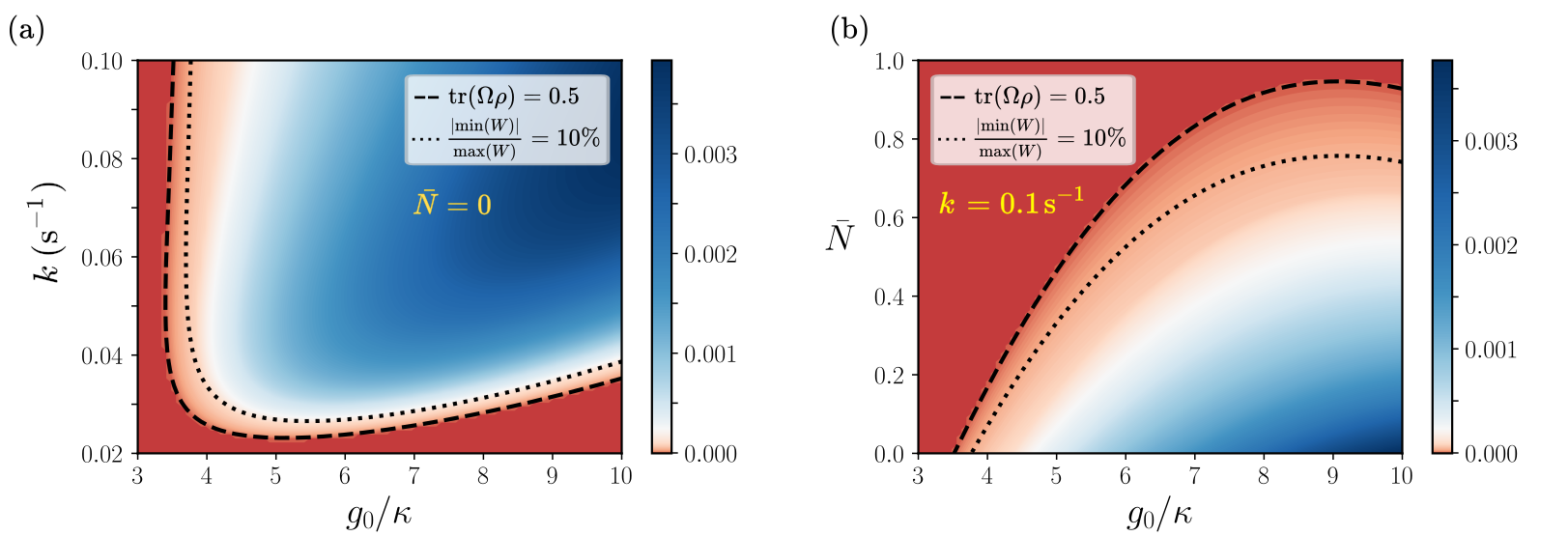}
    \caption{The negative volume indicator $\delta$ of the mechanical steady state generated via continuously-driven nonlinear cavity optomechanics. The parameters used here are the same as in Fig.~\ref{fig_5_main}. (a) The negative volume indicator $\delta$ plotted as a function of the photon flux $k$ and the ratio $g_{0}/\kappa$. Here, $\delta$ shows qualitatively similar behaviour as $|\mathrm{min}(W)|$ from Fig.~\ref{fig_5_main}(b). (b) The negative volume indicator $\delta$ plotted as a function of the occupation of the mechanical thermal environment $\bar{N}$ and the ratio $g_{0}/\kappa$. Again, $\delta$ shows qualitatively similar behaviour as $|\mathrm{min}(W)|$ from Fig.~\ref{fig_5_main}(c).}
    \label{appendix_fig_delta}
\end{figure*}
%%%%%%%%%%%%%%%%%%%%%%%%%%%%%%%%%%%%%%%%%%%%%%%%%%%%%%%%%%%%%%%%%
%%%%%%%%%%%%%%%%%%%%%%%%%%%%%%%%%%%%%%%%%%%%%%%%%%%%%%%%%%%%%%%%%

\end{document}